\documentclass[9pt,a4paper,twocolumn,twoside]{tau-class/tau}
\usepackage[english]{babel}
\usepackage{caption}
\usepackage{bm}
\usepackage{amsmath}
\usepackage{amssymb}
\usepackage{svg}
\usepackage{algorithm}
\usepackage{algpseudocode}
\usepackage{relsize}
\usepackage{cancel}
\usepackage[normalem]{ulem}
\usepackage[section]{placeins}
\algrenewcommand\alglinenumber[1]{}

\algrenewcommand\algorithmicrequire{\textbf{Require:}}
\algrenewcommand\algorithmicensure{\textbf{Ensure:}}
\algrenewcommand\algorithmicforall{\textbf{for all}}
\algrenewcommand\algorithmicif{\textbf{if}}
\algrenewcommand\algorithmicthen{\textbf{then}}
\algrenewcommand\algorithmicelse{\textbf{else}}
\algrenewcommand\algorithmicend{\textbf{end}}
\algrenewcommand\algorithmicfor{\textbf{for}}
\algrenewcommand\algorithmicwhile{\textbf{while}}
\algrenewcommand\algorithmicreturn{\textbf{return}}

\journalname{arXiv Preprint}
\title{Addressing the gravitational collapse of a massless scalar field with Physics-Informed Neural Networks}

\author[1]{* Antonio Ferrer-Sánchez}
\author[1,3]{Nino Villanueva-Espinosa}
\author[1,2]{Carlos Hernani Morales}
\author[5]{Roberto Ruiz de Austri-Bazan}
\author[3,4]{José A. Font}
\author[1,6]{José David Martín-Guerrero}
\author[7]{Matthew W. Choptuik}

\affil[1]{Intelligent Data Analysis Laboratory (IDAL), Department of Electronic
Engineering, ETSE-UV, University of Valencia, Spain.}
\affil[2]{Quantum Spain, 46100 Burjassot, Valencia, Spain.}
\affil[3]{Departamento de Astronomía y Astrofísica, Universitat de València, Dr. Moliner 50, Burjassot, 46100, Valencia, Spain.}
\affil[4]{Observatori Astronòmic, Universitat de València, Catedrático José Beltrán 2, Paterna, 46980, Valencia, Spain.}
\affil[5]{Instituto de Física Corpuscular CSIC-UV, c/Catedrático José Beltrán 2, Paterna, 46980, Valencia, Spain.}
\affil[6]{Valencian Graduate School and Research Network of Artificial Intelligence (ValgrAI), Spain.}
\affil[7]{Department of Physics and Astronomy, University of British Columbia, Vancouver BC, V6T 1Z1, Canada.}

\professor{*e-mail: \href{mailto:antonio.ferrer-sanchez@uv.es}{Antonio.Ferrer-Sanchez@uv.es}}

\institution{Department of Electronic Engineering, ETSE-UV, University of Valencia.}
\footinfo{arXiv Preprint}
\leadauthor{Antonio Ferrer-Sanchez et al.}

\begin{abstract}    
The gravitational collapse of a massless scalar field remains a demanding benchmark for numerical methods in numerical relativity, as it exhibits critical behavior at the boundary between dispersion and black hole formation. In this work we revisit this problem by relying on Physics-Informed Neural Networks (PINNs) as flexible solvers for partial differential equations, thereby providing a comparative assessment of several recent neural architectures. Building on the Einstein-massless-Klein-Gordon formulation in polar-areal coordinates, we consider four initial-value problems encompassing subcritical, critical, and supercritical regimes and use high-resolution finite-difference simulations as reference solutions. Our study is primarily comparative: we evaluate several state-of-the-art deep learning architectures, including vanilla and high-precision PINNs, sinusoidal-feature and quadratic-residual variants, and Kolmogorov-Arnold Networks, all trained under a common loss design that encodes the field equations, boundary conditions, and causal time-space enforcement, together with a novel adaptive spacetime sampling. Within this framework we also introduce ModPINN, a modest modification of standard PINNs that augments standard multilayer perceptrons with coordinate embeddings, quadratic layers, and other common ingredients in recent literature. This  study shows that deep-learning-based methods can reproduce finite-difference solutions for the scalar field and the spacetime metric with competitive accuracy using significantly fewer collocation points than more traditional  methodologies. While no single architecture dominates in all regimes, ModPINN achieves particularly stable and accurate solutions near criticality, indicating that suitably designed embeddings and adaptive sampling can enhance the robustness of PINNs for challenging gravitational-collapse scenarios.
\end{abstract}


\begin{document}
		
    \maketitle 
    \thispagestyle{firststyle} 
    \tauabstract 
    
\section{Introduction}
\label{sec:Introduction}

\subsection{Problem setup and basic equations}

The study of gravitational collapse and black hole formation stands as one of the most interesting and revealing problems in modern physics. Beyond its astrophysical relevance, gravitational collapse provides an exceptional framework for testing the nonlinear, strong-field dynamics of general relativity (GR) and fundamental conjectures such as cosmic censorship. A particularly insightful and motivating aspect of this field arises in the so-called \textit{critical collapse} problem, first discovered by Choptuik in the early 1990s~\cite{Choptuik_seminal}, where the threshold between dispersion and black hole formation in the collapse of a massless scalar field was found to display a universal and self-similar behavior. 

Within the framework of GR, one can consider a spherically symmetric massless scalar field $\phi$ minimally coupled to the spacetime metric $g_{\mu\nu}$. By characterizing the initial data of this system in terms of a one-parameter family labeled by $p$, Choptuik showed the existence of a critical value $p^{*}$ that sharply separates solutions which disperse to flat spacetime ($p<p^{*}$) from those that form a black hole ($p>p^{*}$). 

Near this threshold, the dynamics exhibit critical phenomena: (i) universality, in the sense that different families of initial data approaching $p^{*}$ lead to similar universal behavior, largely independent of their detailed profile; (ii) a power-law scaling relation between the mass of the resulting black hole and the distance to the critical parameter, $M_{\mathrm{BH}}\sim\left|p-p^{*}\right|^{\gamma}$, where $\gamma\approx 0.37$ is a universal critical exponent; and (iii) discrete self-similarity (or ``echoing''), whereby the critical solution reproduces itself on progressively smaller spacetime scales with a characteristic logarithmic periodicity. 

These findings reveal that the dynamics in the proximity of the black hole threshold can be characterized by attractor solutions of codimension one in the phase space of GR. In this regime, spacetime curvature can grow without bound outside the event horizon, providing a natural path towards understanding how GR itself approaches the limits of predictability and potentially signaling where quantum gravitational effects must emerge. Consequently, the study of critical collapse helps to deepen our understanding of the nonlinear structure of Einstein's equations and builds direct relations to fundamental issues in high-energy physics, quantum gravity, and the nature of singularity formation, thereby paving the way for both analytical and numerical methods in modern gravitational theory.

The parameter $p$ corresponds to a specific characteristic of the initial distribution of the scalar field, $\phi(t=0,r):=\phi_{0}(r)$, where $t$ and $r$ denote the time and radial components, respectively, in polar-areal coordinates, $\{t,r,\varphi,\theta\}$. These coordinates help to express the time-dependent Schwarzschild-like metric

\begin{equation}
ds^{2}=-\alpha^{2}(t,r)\:dt^{2}+a^{2}(t,r)\:dr^{2}+r^{2}\:d\Omega^{2},
\label{eq:Sch_metric}
\end{equation}
where $d\Omega^{2}=d\theta^{2}+\sin^{2}\theta\:d\varphi^{2}$ is the metric on the unit 2-sphere, $\alpha(t,r)$ represents the lapse function encoding the proper time evolution and $a(t,r)$ stands for the spatial metric component in the radial direction.

The parameter $p$ naturally divides the entire set of initial conditions into two distinct categories: those that lead to the formation of a black hole during evolution and those that do not. Therefore, we can introduce the concept of a ``threshold'' when discussing the formation of an event horizon. Spacetimes evolving from initial conditions near this threshold exhibit properties that, from a mathematical perspective, are analogous to a critical phase transition in statistical mechanics (interested readers are referred to Sections 2.2 and 3.3 in~\cite{Gundlach2007}). In this manner, the parameter $p$ has a critical value $p^{*}$ separating ``subcritical'' solutions for which the mass dissipates ($p<p^{*}$) from ``supercritical'' solutions for which an event horizon forms ($p>p^{*}$), leading to the emergence of an apparent horizon. 
 
Near the critical region, $p\sim p^{*}$, any geometric quantity $X$ characterizing the black hole at the time of its formation---particularly its mass---follows a power law~\cite{PhysRevD.84.044034},
\begin{equation}
\mathrm{ln}\:X=\gamma\:\mathrm{ln}\left|p-p^{*}\right|+f\left(\mathrm{ln}\:\left|p-p^{*}\right|\right),
\label{eq:BH_mass_relation_2}
\end{equation}
where $\gamma$ is a universal scaling exponent independent of both the type of initial data considered and the physical quantity $X$ being measured. Moreover, $f$ stands for a function that can be directly related to a nonuniversal constant, $c_{f}$, which depends on the shape of the initial field $\phi_{0}(r)$ for the particular case of the black hole mass $M_{\mathrm{BH}}$. In this case, relation~(\ref{eq:BH_mass_relation_2}) reads
\begin{equation}
\mathrm{ln}\:M_{\mathrm{BH}}=\gamma\:\mathrm{ln}\left|\Delta p^{*}\right|+\mathrm{ln}\:c_{f}\,,
\label{eq:BH_mass_relation_3}
\end{equation}
highlighting the proportionality between the logarithm of $M_{\mathrm{BH}}$ (or, equivalently, the event horizon radius via $r_{\mathrm{BH}}=2M_{\mathrm{BH}}$) and the logarithm of the deviation of the relevant physical parameter from its critical value, i.e., $\left|\Delta p^{*}\right|:=\left|p-p^{*}\right|$.

For a massless scalar field $\phi(t,r)$ minimally coupled to the metric~(\ref{eq:Sch_metric}), the Einstein tensor is given by
\begin{equation}
G_{ab}=8\pi\left(\nabla_{a}\phi\nabla_{b}\phi-\frac{1}{2}g_{ab}\nabla_{c}\phi\nabla^{c}\phi\right).
\label{eq:Einstein_tensor}
\end{equation}
Together with the mass conservation equation $\nabla_{a}\nabla^{a}\phi=0$, the Einstein-massless-Klein-Gordon (EMKG) system in spherical symmetry, written in terms of the polar-radial coordinates $(t,r)$ and using geometric units ($G=c=1$), reads:

\begin{tauenv}[frametitle=EMKG field equations]
\begin{equation}
\frac{\partial\Phi}{\partial t}=\frac{\partial}{\partial r}\left(\frac{\alpha}{a}\Pi\right),\qquad\frac{\partial\Pi}{\partial t}=\frac{1}{r^{2}}\frac{\partial}{\partial r}\left(r^{2}\frac{\alpha}{a}\Phi\right).
\label{eq:constraint_1}
\end{equation}

\begin{equation}
\frac{\partial\:\mathrm{ln}\:\alpha}{\partial r}-\frac{\partial\:\mathrm{ln}\:a}{\partial r}+\frac{1-a^{2}}{r}=0.
\label{eq:constraint_2}
\end{equation}

\begin{equation}
\frac{\partial\:\mathrm{ln}\:a}{\partial r}-\frac{1-a^{2}}{2r}-\frac{a^{2}}{r}\left(X^{2}+Y^{2}\right)=0.
\label{eq:constraint_3}
\end{equation}
\end{tauenv}

At this stage, one can identify the ``primitive'' set of variables, consisting of $(\phi,\alpha,a)(t,r)$, where the physically admissible ranges are $\alpha\in[0,1]$ and $a\in(1,+\infty)$, while the values of $\phi$ can be considered numerically unbounded. In addition, some auxiliary variables have been introduced.

\begin{equation}
\Phi(t,r)=\frac{\partial\phi}{\partial r}(t,r),\qquad\Pi(t,r)=\frac{a}{\alpha}\frac{\partial\phi}{\partial t}(t,r).
\label{eq:aux_variables_1}
\end{equation}

\begin{equation}
X(t,r)=\sqrt{2\pi}\frac{r}{a}\Phi(t,r),\qquad Y(t,r)=\sqrt{2\pi}\frac{r}{a}\Pi(t,r).
\label{eq:aux_variables_2}
\end{equation}

The set $(\Phi,\Pi)(t,r)$ in~(\ref{eq:aux_variables_1}) is directly associated with the spatial and temporal derivatives of the scalar field, respectively. In terms of these variables, the set $(X,Y)(t,r)$ in~(\ref{eq:aux_variables_2}) can be defined, leading to some relevant mathematical relations involving the derivatives of the mass aspect function, $m(t,r)$, 
\begin{equation}
\frac{\partial m}{\partial r}=X^{2}+Y^{2}, 
\quad\frac{\partial m}{\partial t}=\left(\frac{a}{\sqrt{1-a^{2}}}\right)^{3}\alpha\:X\:Y.
\label{eq:interesting_relation_1}
\end{equation}
For completeness, the spacetime curvature scalar $R$ can also be written in terms of this pair of variables as
\begin{equation}
R=\frac{4}{r^{2}}(X^{2}-Y^{2}).
\label{eq:interesting_relation_2}
\end{equation}

The mass function represents the amount of mass enclosed within a radius $r$ at a given time $t$ during the evolution. It is defined in analogy to the mass in the static Schwarzschild metric in terms of $a(t,r)$,
\begin{equation}
m(t,r)=\frac{r}{2}\left(1-\frac{1}{a^{2}}\right).
\label{eq:mass_aspect}
\end{equation}

In order to determine whether a given initial data setup has resulted in gravitational collapse---and consequently, the formation of an apparent horizon---it is necessary to evaluate the ratio ${2m}/{r}$, commonly referred to as the compactness parameter (or, alternatively, the dimensionless mass aspect). We will denote this parameter by $C$, which can be easily defined in terms of the radial metric coefficient,
\begin{equation}
C(t,r):=\frac{2m(t,r)}{r}=1-\frac{1}{a^{2}(t,r)}.
\label{eq:defining_C}
\end{equation}

Within this formulation, this parameter satisfies $C\in[0,1)$, where the limit $C\to 1$ characterizes the onset of gravitational collapse. From a computational perspective, the formation of an apparent horizon in numerical simulations would be identified by imposing a threshold $\epsilon$ such that $1-C\leq\epsilon$. In this manner, classical numerical methods based on finite-difference methods (FDM) can effectively solve the field equations in both dispersive scenarios of $\phi(t,r)$ and cases involving gravitational collapse (e.g. see~\cite{Choptuik_1998}). Despite the coupling between the different variables, the dynamical evolution is governed by~(\ref{eq:constraint_1}), while~(\ref{eq:constraint_2}) and (\ref{eq:constraint_3}) serve as metric constraints.

\subsection{Deep-learning-based methodologies}

Numerical solutions of the field equations (\ref{eq:constraint_1}-\ref{eq:constraint_3}) have been reported since the seminal work of~\cite{Choptuik_seminal}. Several studies (see \cite{PhysRevD.84.044034,Garfinkle_1999,Sorkin_2005}) focused on investigating critical behavior within generalizations of this framework to spacetimes of higher dimensions. Other extensions took into account massive fields~\cite{Erik_2022}, configurations involving multiple scalar fields~\cite{Gundlach_2019}, or scenarios incorporating more sophisticated or different metric structures, yielding interesting results, such as those reported in~\cite{Benitez_2020} or~\cite{Cai_2017}. 

Despite the many numerical studies, it is worth noting that the most common methodology to solve the equations has been based on FDM. Only recent works have considered alternative methods to FDM to investigate gravitational collapse. In particular, the work of~\cite{finzi_2023} introduced a neural network-based solver that mitigates artificial neural network (ANN) ill-conditioning while exhibiting linear scalability with respect to the number of parameters. Other studies, such as~\cite{hatefi_2023}, developed a model also based on ANNs to investigate critical phenomena in Einstein-axion-dilaton models, finding no evidence of self-similar behavior. On the other hand, \cite{hatefi_2023_2} and~\cite{hatefi_2024} employ Bayesian deep learning models to study the axion-dilaton system, utilizing Sequential Monte Carlo (SMC) sampling and related algorithms to obtain direct estimates of the  critical exponent $\gamma$.

All of these recent studies establish a foundation for the incorporation of deep learning (DL) techniques in the investigation of gravitational collapse scenarios. The present work proposes the adoption of novel methodologies based on ANNs tailored to solve systems of partial differential equations (PDEs) and extract their underlying physics. Since their initial appearance in the scientific literature~\cite{seminal_pinns}, Physics-Informed Neural Networks (PINNs) have emerged as a robust alternative for solving numerical problems governed by PDEs. These methods leverage the universal approximation theorem of ANNs~\cite{Pinkus_1999} to minimize the governing physical equations, incorporating them into the loss function used to train the network parameters. This function is typically augmented with physical constraints---commonly referred to as ``inductive biases''---such as initial and boundary conditions, additional physical information to be minimized, or even agreement with experimental data. Therefore, our primary objective will be to demonstrate that the PINN methodology is fully capable of solving the EMKG equations~(\ref{eq:constraint_1}-\ref{eq:constraint_3}), effectively addressing both the field dispersion and the gravitational collapse scenarios. This approach is proposed as a viable alternative to traditional numerical methods.

The paper is structured as follows. Section~\ref{sec:literature-review} presents an overview of recent literature on PINNs, to motivate our study. Next, Section~\ref{sec:Methodology} introduces our PINN-based methodology, which consists of a modified version of the standard (vanilla) PINN; this section also details the specific formulation used to minimize the loss function. Section~\ref{sec:Results} presents the results obtained for various initial values of the scalar field, including the critical case, with particular emphasis on evaluating the predictive performance of the model in that regime. Finally,  Section~\ref{sec:Conclusions} presents our conclusions and outlines potential directions for future research.

\section{Review of recent literature}
\label{sec:literature-review}

In general terms, the choice of neural architecture is a critical factor, as the number of trainable parameters must be appropriate for the problem at hand, whatever the specific task may be---whether solving PDEs or addressing another application. Among the aspects that define a neural architecture, several may be highlighted: the activation functions employed; the number of layers (depth) and their size (width); whether residual or skip connections are incorporated between layers; and whether any form of embedding is used, among other considerations. Several studies~\cite{Yifan_2024} conclude, in a range of independent physical problems, that relatively shallow but wide networks often outperform deeper but narrower ones. This reveals a concern that usually appears throughout the literature on multilayer perceptrons (MLPs) applied to physics, motivating more recent developments such as Physics-Informed Residual Adaptive Networks (PirateNets)~\cite{piratenets_2024}, which are introduced as residual architectures that allow the model to start from shallow networks, progressively ``revealing'' additional layers only when they are advantageous in reducing the objective cost function. These networks also employ embeddings of the physical coordinates, mapping them into a high-dimensional feature space using random Fourier features~\cite{Tancik_2020}. However, although these embeddings can mitigate the frequency bias typically observed in PINNs~\cite{Sifan_2021} and facilitate the solution of physical problems involving high-frequency behavior, the selection of the maximum frequency scale and the standard deviation of the Gaussian distribution used to sample the frequencies are critical; if chosen poorly, these become problem-dependent hyperparameters that can hinder performance. Other architectures, such as the sf-PINNs proposed in~\cite{Wong_2024}, apply a mapping of the input coordinates to sinusoidal features, encoding them after the first layer, thus increasing the variability of the input gradient and also helping mitigate the said frequency bias.

Stepping somewhat away from traditional architectures, methodological variants such as the Quadratic Residual Networks (QRes)~\cite{qres_2021} aim to replace the standard linear layer with an alternative that introduces second-order interactions between the trainable parameters and the input, becoming more parameter-efficient and faster to train, in general, on scientific problems. Within the most recent DL literature, Kolmogorov-Arnold Networks (KANs)~\cite{kan_2025} have emerged as an alternative to conventional MLPs, which have also been applied to PINNs. These mainly consist of replacing the trainable weights of linear layers with learnable univariate functions located at the nodes, yielding high expressivity with substantially fewer parameters. Finally, architectures such as the PINNsFormer~\cite{pinnsformer_2024} implement a transformer-based methodology for time-dependent data, translating each point $(t,r)$ into a short temporal sequence of $k$ steps that is fed into a standard encoder-decoder architecture, thus learning and enforcing temporal dependencies. In contrast, although PINNMamba~\cite{mamba_2025} shares certain similarities with the preceding approach, it replaces the MLP backbone with an encoder-only Mamba~\cite{ssm_2024} (selective state-space) network that models short, overlapping temporal subsequences. This design yields a continuous solution over the time of the physical problem, with consistency across overlaps enforced via a contrastive loss between the subsequences. All of these architectural approaches present distinct strengths and limitations, and each has been evaluated on a broad range of physics problems in their respective references. In our case, in Section~\ref{sec:Results}, we undertake a comparative assessment of their performance in solving a variety of sub- and supercritical gravitational collapse problems of the massless scalar field minimally coupled to the spacetime metric.

Regarding domain sampling, this step is also crucial to the methodology, since the collocation points---here the spacetime coordinates $(t,r)$---serve as training data for the network. Several strategies for generating these points may be considered, ranging from simple, uniformly spaced grids or randomly generated samples for a particular probability distribution to more sophisticated techniques such as Sobol sequences~\cite{sobol_1967} or Latin hypercube sampling~\cite{stein_1987,mckay_2000}. Other approaches incorporating specialized deterministic distributions can also be considered. For example, methods that employ Chebyshev-distributed nodes in space combined with exponentially spaced temporal points enable more frequent and precise sampling near boundaries or localized features~\cite{vittorio_2025}. From a general perspective, any technique that involves non-uniform local refinement sampling can offer a strategy tailored to the specific problem, although such methods remain relatively uncommon in the PINN literature. Despite the fact that the performance of all these methods may vary depending on the specific PINN problem at hand, all of them share one common feature: they rely on fixed sampling strategies that remain static throughout the entire training process. In a comprehensive review, the authors of~\cite{florido_2024} concluded that there is no universal rule for defining a general and optimal sampling strategy for PINNs, saying that it is necessary to define the sampling in an intelligent manner and tailor the methodology to the particular scenario.

Within the range of adaptive sampling methods, those based on probability distributions deserve special mention, highlighting~\cite{chenxi_2023} where the authors introduced the Residual-based Adaptive Distribution (RAD) which focused on resampling points with a probability proportional to their PDE residual. In studies such as~\cite{nabian_2021}, the authors propose sampling techniques based on the significance assigned by the network to different regions of the training domain, finding that PINNs achieve faster convergence when the points are sampled according to a weighted probability density function (PDF) that depends on the loss. Some extensions of this idea are explored in works such as~\cite{Xiaotian_2024}, where a PDF is defined to continuously concentrate training points in regions with high loss values while still maintaining coverage in areas with lower errors. This balance is crucial, as completely neglecting low-error regions could lead to systematic forgetting by the PINN. Other studies focus more on repositioning the collocation points. For example, the RANG method (Residual-based Adaptive Node Generation) proposed in~\cite{peng_2022} treats the sampling points as variable-density radial basis function (RBF) nodes, allowing adaptive redistribution based on the residual.

\section{Methods}
\label{sec:Methodology}

Incorporation of physical information into neural networks for use as universal approximators of differential equations has been discussed in the literature for several decades~\cite{Lee_1990,Psichogios_1992,Dissanayake_1994}. However, the emergence of computational techniques, such as automatic differentiation~\cite{Baydin_2018}, marked the formal introduction of PINNs as a distinct concept, which were originally formulated as fully connected feedforward neural networks composed of $\mathrm{K}$ hidden stacked layers~\cite{Cuomo_2022}. These networks take a vector $\bm{x}$ as input, producing as output a tensor of target variables which we will denote by $\pmb{\mathcal{F}}(\bm{x};\Theta):=\pmb{\mathcal{F}}_{\Theta}(\bm{x})$, where $\Theta$ represents the set of all trainable parameters of the architecture---typically, for a vanilla model, $\Theta=\{\bm{W}_{k},\bm{b}_{k}\}_{1\leq k\leq K}$, where $\bm{W}_{k}\in\mathbb{R}^{N_{k}\times N_{k}}$ and $\bm{b}_{k}\in\mathbb{R}^{k}$ are the network weights and biases of the $k$-th layer, respectively. Given the requirement of non-linearity, a set of activation functions denoted by $\{\sigma_{k}\}_{1\leq k\leq K-1}$ is applied to the output of each hidden layer\footnote{Note that the set of activation functions spans only up to the penultimate layer. In general, the output layer corresponds to physical variables that typically require independent activation functions to restrict them within the acceptable physical range.}. These activation functions can be implemented either at the layer level or neuron-wise, which can also be trained~\cite{Jagtag_2020,Khademi_2025}.

In the context of the present study, a spacetime domain defined as $(t,r)\in[0,T]\times\Omega$, where $T$ denotes the maximum evolution time, is considered. A vanilla PINN is constructed by minimizing a given differential operator~(\ref{eq:differential_operator}), typically subject to specific constraints that the predicted solution must satisfy, such as initial conditions in time~(\ref{eq:IC_conditions}) and boundary conditions in space~(\ref{eq:BC_conditions}).

\begin{equation}
\mathrm{P}\left(t,r;\frac{\partial\pmb{\mathcal{F}}}{\partial t},\frac{\partial^{2}\pmb{\mathcal{F}}}{\partial t^{2}},\ldots;\frac{\partial\pmb{\mathcal{F}}}{\partial r},\frac{\partial^{2}\pmb{\mathcal{F}}}{\partial r^{2}},\ldots;\frac{\partial^{2}\pmb{\mathcal{F}}}{\partial r\partial t},\ldots\right)=0.
\label{eq:differential_operator}
\end{equation}

\begin{equation}
\mathcal{IC}(t,r)=0,\qquad(t,r)\in\{0\}\times\Omega.
\label{eq:IC_conditions}
\end{equation}

\begin{equation}
\mathcal{BC}(t,r)=0,\qquad(t,r)\in(0,T]\times\partial\Omega.
\label{eq:BC_conditions}
\end{equation}

In general, for the case of vanilla PINNs, the loss function employed, $\mathcal{L}$, given by~(\ref{eq:L_vanilla}) below, involves both the minimization of the governing PDEs, denoted by $\mathcal{L}_{\mathrm{PDE}}$, and the incorporation of the additional constraints, $\mathcal{L}_{i}$, among which the initial and boundary conditions are considered. A weight vector $\vec{\pmb{\lambda}}=\left(\lambda_{\mathrm{PDE}},\lambda_{\mathcal{IC}},\lambda_{\mathcal{BC}},\ldots\right)$ is defined to combine the different components of the loss into a single objective function.
\begin{equation}
\mathcal{L}=\lambda_{\mathrm{PDE}}\mathcal{L}_{\mathrm{PDE}}+\sum_{i}\lambda_{i}\mathcal{L}_{i}.
\label{eq:L_vanilla}
\end{equation}
A quadrature point vector $\vec{\pmb{N}}=\left(N_{\mathrm{PDE}},N_{\mathcal{IC}},N_{\mathcal{BC}},\ldots\right)$ is defined to specify the sampling locations used to evaluate each component. Although alternatives have been studied~\cite{Chuwei_2022}, these terms are typically defined individually as mean squared errors (MSE): in the case of $\mathcal{L}_{\mathcal{IC}}$ and $\mathcal{L}_{\mathcal{BC}}$, as the MSE between the predicted and true values; and in the case of $\mathcal{L}_{\mathrm{PDE}}$, as the MSE of the residuals of the governing equations.

\subsection{Construction of the loss function}

Based on the definitions provided in the previous section and considering the compactness parameter $C$ defined in~(\ref{eq:defining_C}), the evolution equation for the scalar field can be expressed as, 
\begin{equation}
\mathrm{eq}_{1a}:=\frac{\partial\Phi}{\partial t}-\frac{\partial}{\partial r}\left(\alpha\:\Pi\:\sqrt{1-C}\right).
\label{eq:PDE_1a}
\end{equation}
\begin{equation}
\mathrm{eq}_{1b}:=\frac{\partial\Pi}{\partial t}-\frac{1}{r^{2}}\frac{\partial}{\partial r}\left(r^{2}\:\alpha\:\Phi\sqrt{1-C}\right).
\label{eq:PDE_1b}
\end{equation}
On the other hand, the Hamiltonian constraint, which enforces the balance between the gravitational and scalar field energies at each radial coordinate $r$, along with the slicing condition, which stands for the momentum constraint under a zero-shift gauge and serves to fix the lapse function on each time slice, can be reformulated as
\begin{equation}
\mathrm{eq}_{2}:=r(1-C)\frac{\partial\:\mathrm{ln}\:\alpha}{\partial r}-\frac{r}{2}\frac{\partial C}{\partial r}-C.
\label{eq:PDE_2}
\end{equation}
\begin{equation}
\mathrm{eq}_{3}:=\frac{\partial m}{\partial r}-\left(X^{2}+Y^{2}\right),
\label{eq:PDE_3}
\end{equation}
respectively. Considering that the evolution of $\phi(t,r)$ is second-order in time, it is necessary to specify its first temporal derivative as part of the initial data. Although there are several options that can be chosen, for the purpose of the present study we will use ``time-symmetric initial data'', that is,  $\frac{\partial\phi}{\partial t}\Bigl|_{t=0}=0$, or equivalently, $\Pi(0,r)=0$.

\begin{algorithm*}[t]
\caption{Adaptive weighted–CDF remeshing of radial coordinate.}
\begin{algorithmic}[1]
  \Require Spacetime tensor $(t,r)\in\mathbb{R}^{N_{\mathrm{t}}N_{\mathrm{r}} \times 2}$, residuals $\mathcal{L}(t,r)\in\mathbb{R}^{N_{\mathrm{t}}N_{\mathrm{r}}}$, adaptivity parameter $\Lambda > 0$, domain bounds $[r_{\min}, r_{\max}]$.
  \Ensure Updated $(t,r^{\mathrm{new}})$ with remeshed radial coordinates.

  \State \textbf{1. Reshape input:} Reshape $r\to r_{\mathrm{old}}[n,i]$, $\mathcal{L}\to\mathcal{L}[n,i]$ of shape $N_{\mathrm{t}}\times N_{\mathrm{r}}$.
  \State \textbf{2. Sort each time slice:} Sort $r_{\mathrm{old}}[n,\cdot]$ along $r$ and permute $\mathcal{L}[n,\cdot]$ accordingly to obtain $\left(r_{n,j}, \mathcal{L}_{n,j}\right)$.
  \State \textbf{3. Normalize residuals:} Linearly map $\mathcal{L}_{n,j}$ to $[0,1]$ as
  \[
    \mathcal{L}^{\mathrm{norm}}_{n,j} = \frac{\mathcal{L}_{n,j}-\min_j \mathcal{L}_{n,j}}{\max\left(\max_j \mathcal{L}_{n,j}-\min_j \mathcal{L}_{n,j},\varepsilon\right)}.
  \]
  \State \textbf{4. Compute density:} Define $m_{n,j} = 1 + \Lambda \mathcal{L}^{\mathrm{norm}}_{n,j}$.
  \State \textbf{5. Compute weights:} For each interval,
  \[
    \Delta r_{n,j} = r_{n,j+1} - r_{n,j}, \quad w_{n,j} = \frac{1}{2}(m_{n,j} + m_{n,j+1}) \Delta r_{n,j}.
  \]
  \State \textbf{6. Build cumulative distribution:} Let $W_{n,0} = 0$ and compute cumulative sums: $W_{n,k} = \sum_{j=0}^{k-1} w_{n,j}$.
  \State \textbf{7. Define uniform target masses:} For each $i$,
  \[
    c_{n,i}=\frac{i}{N_r-1}W^{\mathrm{tot}}_n,\quad \mathrm{where}\quad W^{\mathrm{tot}}_n=W_{n,N_r-1}.
  \]
  \State \textbf{8. Invert weighted–CDF:} For each $c_{n,i}$, find $j$ such that $W_{n,j}\leq c_{n,i}<W_{n,j+1}$ and interpolate:
  \[
    \beta=\frac{c_{n,i}-W_{n,j}}{w_{n,j}},\quad r^{\mathrm{new}}_{n,i}=r_{n,j}+\beta\Delta r_{n,j}.
  \]
  \State Clamp $\beta$ to $[0,1]$.
  \State \textbf{9. Handle endpoints:} Set $r^{\mathrm{new}}_{n,0}=r_{n,0}$ and $r^{\mathrm{new}}_{n,N_r-1}=r_{n,N_r-1}$.
  \State If $W^{\mathrm{tot}}_n\leq 0$, mark slice as degenerate and set $r^{\mathrm{new}}_{n,j}=r_{n,j}$ for all $j$.
  \State \textbf{10. Restore original order:} Unsort $r^{\text{new}}_{n,i}$ to match the original index order.
  \State \textbf{11. Clamp to domain:} Enforce $r^{\text{new}}_{n,i}\in[r_{\min},r_{\max}]$.
  \State \textbf{12. Preserve initial slice:} For $n = 0$, set $r^{\text{new}}_{0,i} \coloneqq r_{\text{old}}[0,i]$.
  \State \textbf{13. Reassemble tensor:} Overwrite the radial coordinates in the original $(t,r)$ spacetime with the new set $\{r^{\text{new}}_{n,i}\}$ once flattened.
\end{algorithmic}
\label{algo:algorithm_1}
\end{algorithm*}

In our methodology, a two-dimensional spacetime domain is numerically discretized as $(t,r)\in[0,T]\times[0,R]$, where the parameters $T$ and $R$ denote the maximum values reached in the numerical simulation and must be chosen sufficiently large to encompass all relevant physical information. The resulting grid thus comprises $N_{\mathrm{t}}$ points in the temporal dimension and $N_{\mathrm{r}}$ in the radial direction. Therefore, the original loss function depends on both time and space, since a loss value is computed at each collocation point. In that sense, the first three terms in the loss equation
\begin{equation}
\mathcal{L}(t,r)=
\begin{cases}
\displaystyle
\lambda_{\mathcal{IC}}\:\bigl\lvert \pmb{\mathcal{F}}_{\Theta}(0,r)-\pmb{\mathcal{F}}_{\mathrm{GT}}(0,r)\bigr\rvert^2,
& (t,r)\in\{0\}\times[0,R],\\[1ex]
\displaystyle
\lambda_{\mathcal{BC},\:\mathrm{left}}\:\bigl\lvert \pmb{\mathcal{F}}_{\Theta}(t,0)-\pmb{\mathcal{F}}_{\mathrm{GT}}(t,0)\bigr\rvert^2,
& (t,r)\in[0,T]\times\{0\},\\[1ex]
\displaystyle
\lambda_{\mathcal{BC},\:\mathrm{right}}\:\left(\bigl\lvert \mathcal{R}_{\phi}\bigr\rvert^2+\bigl\lvert \mathcal{R}_{\alpha}\bigr\rvert^2+\bigl\lvert \mathcal{R}_{C}\bigr\rvert^2\right),
& (t,r)\in[0,T]\times\{R\},\\[1ex]
\displaystyle
\sum_{i}\lambda_{i}\:\lvert\mathrm{eq}_i(t,r)\rvert^2,
& (t,r)\in(0,T]\times(0,R].
\end{cases}
\label{eq:Loss_vs_t_r}
\end{equation}
highlight the conditions that need to be treated separately: the initial temporal slice ($t=0$), which corresponds to the initial conditions, and the first and last spatial slices ($r=0$ and $r=R$), corresponding to the boundary conditions. In this context, the subscript ``GT'' refers to ``ground truth'', referring to some baseline numerical solution that is being used. The sets of conditions are computed a priori for the three variables of interest, $\pmb{\mathcal{F}}_{\mathrm{GT}}=\left(\phi,\alpha,C\right)_{\mathrm{GT}}$, imposing $t=0$ and $r=0$ in the field equations~(\ref{eq:PDE_1a}--\ref{eq:PDE_3}) and solving them directly using FDM implemented in modern numerical frameworks~\cite{scipy_2020}. In addition, for points sufficiently far from the source---that is, the limit $r\gg 0$ (or $r\to R$)---we impose the standard $1/r$ falloffs of the metric and field variables to ensure asymptotic flatness,
\begin{equation}
\mathcal{R}_{\phi}=\frac{\partial}{\partial r}\left[r\phi\right],\quad\mathcal{R}_{\alpha}=\frac{\partial}{\partial r}\left[r(\alpha-1)\right],\quad\mathcal{R}_{C}=\frac{\partial}{\partial r}\left[rC\right].
\label{eq:falloff_conditions}
\end{equation}
The enforcement of restrictions $\left|\mathcal{R}_{\alpha}\right|^{2},\left|\mathcal{R}_{C}\right|^{2}\to 0$ yields the asymptotics $C\sim\frac{2M_{\mathrm{ADM}}}{r}$ and $\alpha=\sqrt{1-\frac{2M_{\mathrm{ADM}}}{r}}=1-\frac{M_{\mathrm{ADM}}}{r}+\mathcal{O}(r^{-2})$, the latter being the lapse for the exterior Schwarzschild metric and where $M_{\mathrm{ADM}}$ represents the time-conserved Arnowitt-Deser-Misner mass. Finally, demanding $\frac{\partial}{\partial r}\left(r\phi\right)=0$ yields $\phi\sim\frac{\mathrm{const.}}{r}$, the expected dispersal of a massless scalar. These conditions guarantee that the solution given by the network approaches the Schwarzschild asymptotics of an asymptotically flat spacetime, tying the falloff of $\alpha$ and $C$ directly to the ADM mass.


Regarding the resolution of PDEs, the final term in~(\ref{eq:Loss_vs_t_r}) consists of several terms, each representing the squared modulus of a given equation. On the other hand, the set $\{\lambda_{i}\}$ represents the weights assigned to the different equations. In addition, $\lambda_{\mathcal{IC}}$ denotes the coefficient associated with the initial conditions over time, while the pair $(\lambda_{\mathcal{BC},\:\mathrm{left}},\lambda_{\mathcal{BC},\:\mathrm{right}})$ specifies the weights assigned to the boundary conditions at $r=0$ and $r=R$, respectively.

This construction yields a quadrature tensor for the loss function, $\mathcal{L}(t_{n},r_{i})$, evaluated at each collocation point. To advance further, one may adopt the approach proposed by~\cite{Wang_2024}, who proved that PINNs tend to exhibit biased learning solutions at later times prematurely, often causing the model to converge to failure modes. Consequently, for a given time value $t_{n}$, a weighting factor $\omega_{t}(t_{n})\in[0,1]$ is introduced, 
\begin{equation}
\mathcal{L}=\frac{1}{N_{\mathrm{t}}N_{\mathrm{r}}}\sum_{n=0}^{N_{\mathrm{t}}}\omega_{t}\left(t_{n}\right)\sum_{i=0}^{N_{\mathrm{r}}}\omega_{r}\left(t_{n},r_{i}\right)\:\mathcal{L}\left(t_{n},r_{i}\right).
\label{eq:Loss}
\end{equation}

\begin{equation}
\omega_{t}(t_{n})=\mathrm{exp}\left(-\epsilon_{\mathrm{t}}\sum_{m=0}^{n-1}\mathcal{L}\left(t_{m}\right)\right).
\label{eq:omega_t}
\end{equation}

This weight is computed using an exponential function of the cumulative sum of all loss values evaluated at earlier time steps, $\mathcal{L}\left(t_{m<n}\right)$, modulated by a hyperparameter $\epsilon_{\mathrm{t}}$, which governs the strength of this causality enforcement mechanism. Within the framework of the original method, the time-dependent loss function $\mathcal{L}(t_{n})$ is obtained by averaging (or, alternatively, summing) over the spatial dimension. In our approach, we propose an analogous strategy for handling the radial dimension by introducing an additional weighting factor $\omega_{r}(t_{n},r_{i})\in[0,1]$,
\begin{equation}
\omega_{r}(t_{n},r_{i})=\mathrm{exp}\left(-\epsilon_{\mathrm{r}}\sum_{j=0}^{i-1}\mathcal{L}\left(t_{n},r_{j}\right)\right).
\label{eq:omega_r}
\end{equation}
This weight allows sequential learning along the spatial coordinate for a fixed temporal slice, progressively unlocking spatial points in an iterative manner as the previous losses $\mathcal{L}\left(t_{n},r_{j<i}\right)$ are minimized. Similarly, a new hyperparameter $\epsilon_{\mathrm{r}}$ is introduced to regulate the strength of this \textit{spatial enforcement} mechanism. In general, $\epsilon_{\mathrm{r}}$ will be different from its temporal counterpart $\epsilon_{\mathrm{t}}$, allowing independent treatment of the learning dynamics in both dimensions.

The use of a loss function of the form~(\ref{eq:Loss})-(\ref{eq:omega_r}) enables the network to perform sequential learning, mimicking the behavior of traditional FDM. In this way, learning unfolds progressively through the training process, gradually opening the learning space to the model and preventing the network from ``getting ahead'' by trying to learn regions that are posterior to those that have not yet been properly minimized. However, despite its clear advantages during training, enforcing sequentiality in learning can also present some drawbacks, including increased computational time and potential issues such as the cost gets stuck in a specific subset of points $(t,r)$ if the hyperparameter pair $(\epsilon_{\mathrm{t}},\epsilon_{\mathrm{r}})$ is set too high. Therefore, a thorough sensitivity analysis of the network is necessary with respect to these parameters.

\subsection{Adaptive spacetime sampling}

\begin{figure}[h]
    \centering
    \includegraphics[width=0.95\columnwidth]{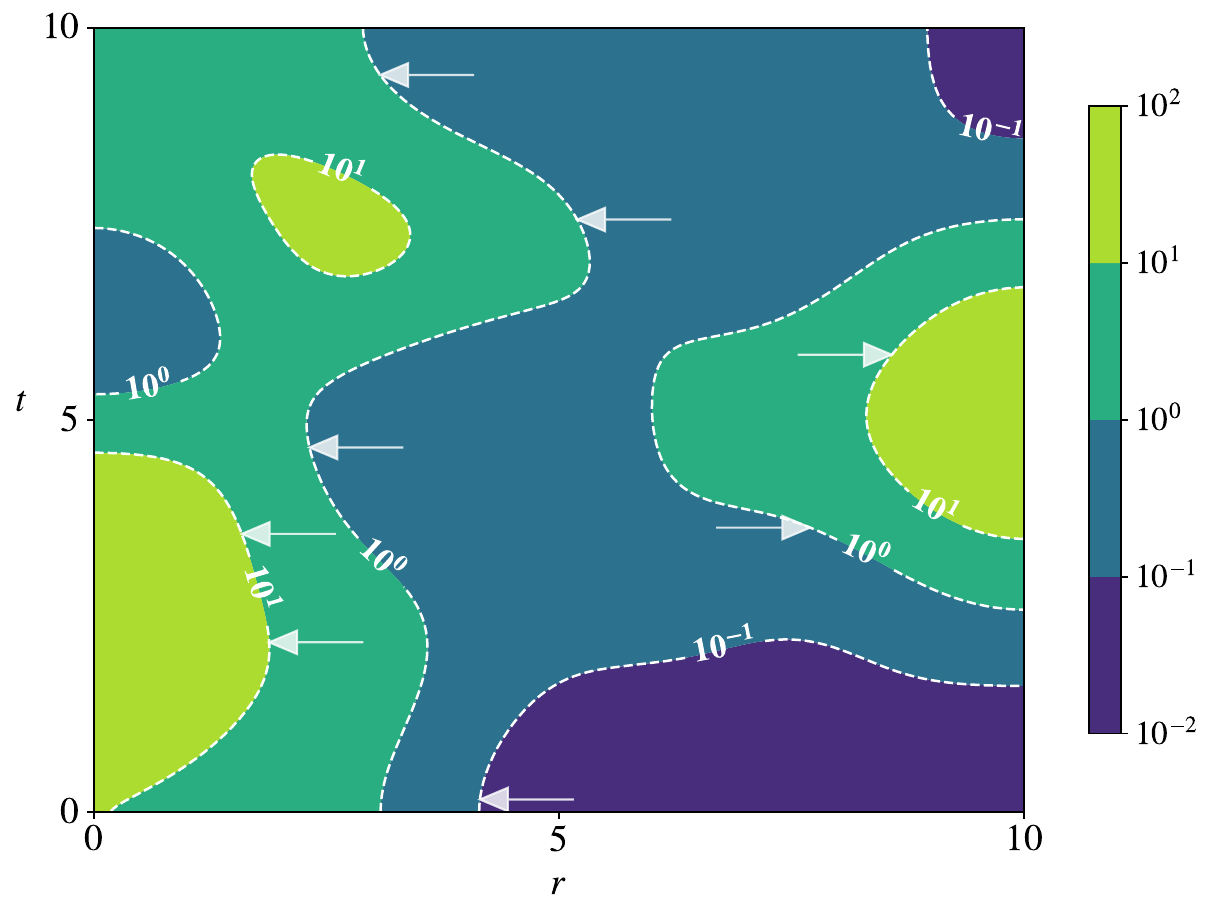}
    \caption{Diagrammatic representation of the adaptive model. The figure shows the landscape of a sample loss function with respect to the $(t,r)$ sampling, whose values span several orders of magnitude. The arrows indicate the direction of movement of the collocation points, which intuitively shift towards regions having higher PDE residuals.}
    \label{fig:adaptive_sampling}
\end{figure}

\noindent Here we propose a method for redistributing the collocation points along the radial dimension $r$ of the spacetime domain considered (see visual aid in figure~\ref{fig:adaptive_sampling}). This approach is guided by the performance of the model at each point, using a weighted cumulative distribution function (CDF). Starting from a two-dimensional sampling in time and space, $(t,r)$, we develop an algorithm that adaptively shifts the spatial points for each fixed temporal slice. Specifically, for a given time $n$, the algorithm starts by computing the normalized error (residual) as 
\begin{equation}
\mathcal{L}_{\mathrm{norm}}(t_{n},r_{i})=\frac{\mathcal{L}\left(t_{n},r_{i}\right)-\mathrm{min}\:\mathcal{L}\left(t_{n},r_{i}\right)}{\mathrm{max}\:\mathcal{L}\left(t_{n},r_{i}\right)-\mathrm{min}\:\mathcal{L}\left(t_{n},r_{i}\right)}\in[0,1],
\label{eq:loss_norm}
\end{equation}
where the loss is calculated using~(\ref{eq:Loss_vs_t_r}).

Points within the domain can be interpreted as ``particles'' subjected to a virtual ``force'' that drives them from their original locations to new positions where the model can utilize them more effectively. To formalize this behavior, a target density function is defined, 
\begin{equation}
m\left(t_{n},r_{i}\right)=1+\Lambda\:\mathcal{L}_{\mathrm{norm}}\left(t_{n},r_{i}\right),
\label{eq:m_density}
\end{equation}
which depends on the normalized loss and a hyperparameter $\Lambda$ that regulates how aggressive the displacement of the points is.

With this framework in place, the core idea is to use the numerically approximated cumulative integral of the target density to determine the new collocation positions. First, we compute the weights between adjacent spatial locations as the average of the density values at two consecutive spatial points for a given time step, as 
\begin{equation}
w\left(t_{n},r_{j}\right)=\frac{m\left(t_{n},r_{j}\right)+m\left(t_{n},r_{j+1}\right)}{2}\:\left(r_{j+1}-r_{j}\right).
\label{eq:w_density}
\end{equation}

These weights are used to approximate the area under the density curve between each pair of points, allowing for the computation of the cumulative ``mass'' via
\begin{equation}
W\left(t_{n},r_{k}\right)=\sum_{j<k}w\left(t_{n},r_{j}\right).
\label{eq:W}
\end{equation}

Target mass values are then selected to be evenly spaced between the original and the total accumulated mass $W_{n}^{\mathrm{tot}}$. By inverting the CDF, we obtain a new set of locations that satisfy this distribution, except at $(t,r)\in[0,T]\times\{0\}$ and $(t,r)\in[0,T]\times\{R\}$, where initial and final boundary conditions are enforced. In addition, the spatial positions at $t=0$ are also kept fixed so as not to affect the calculation of the temporal initial constraints. The entire process is described in detail, step by step, in Algorithm~\ref{algo:algorithm_1}.

\subsection{Model architecture}

\begin{figure*}[t]
    \centering
    \includegraphics[width=2.0\columnwidth]{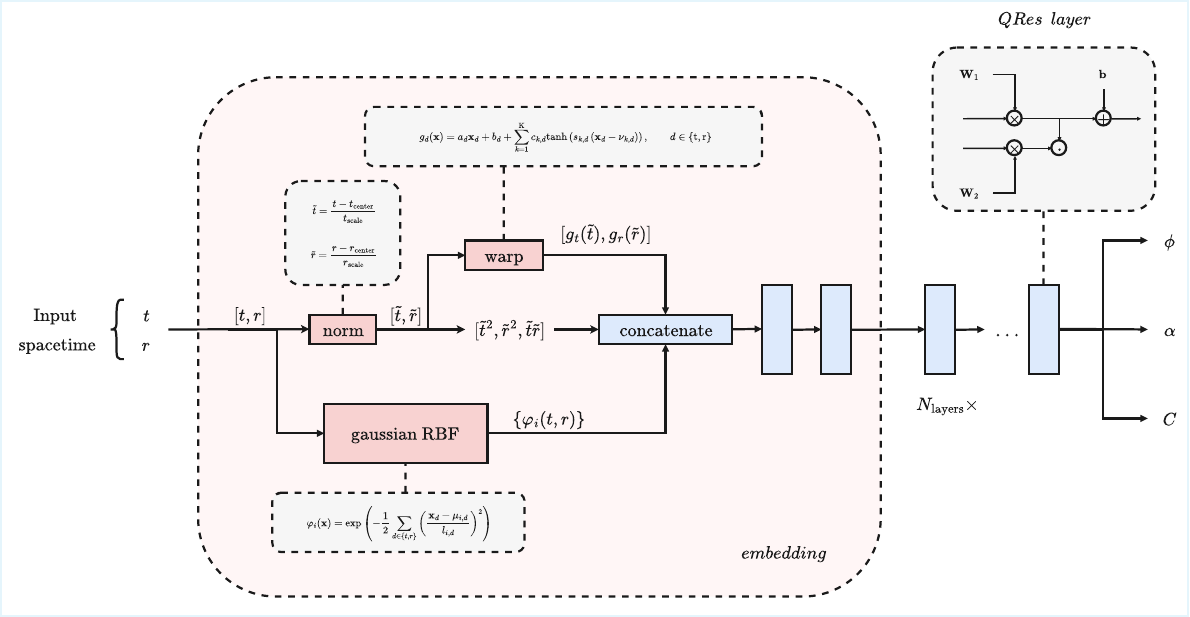}
    \caption{The proposed methodology for the ModPINN. The embedding block of the model modulates $(t,r)$ via a hyperbolic-tangent ``warp'', augments them with low-order polynomials of the normalized coordinates together with a dictionary of $\mathrm{M}$ Gaussian transforms, and projects the concatenated features through stacked quadratic (QRes) layers to output the physical set $(\phi,\alpha,C)$.}
    \label{fig:methods}
\end{figure*}

Throughout this work, several novel methodologies recently introduced in the literature have been considered with the aim of establishing a comparative analysis among them. At the same time, an attempt has been made to introduce a slightly adapted approach of our own, tailored to the specific physical problem under study. This neural architecture is illustrated in the diagram of figure~\ref{fig:methods}, and, for the sake of simplicity in terminology, we shall refer to it as ModPINN---as an alias for ``modified PINN''---throughout the remainder of the text. In general terms, this and any other physics-informed model presented in this work take as input the spacetime coordinates $(t,r)$, process them according to the chosen architecture, and output the set of physical variables, $\left(\phi,\alpha,C\right)_{\Theta}$. These variables are then used to compute the PDEs and, consequently, the loss metric. Although most of the physics of the problem is encapsulated in the proposed cost function~(\ref{eq:Loss}), the neural architecture has also been shown to be a fundamental component in terms of performance, as it can usually be informed by physical principles as well. Consequently, it is worth highlighting the main aspects of our proposed methodology.

\textbf{Embedding.} One of the most well-known issues affecting PINNs is their spectral bias, which often leads to higher frequencies being either unresolved or incorrectly captured~\cite{Sifan_2021,Rahaman_2019}. Since many physical problems involve multiscale solutions or sharp gradients, understanding and mitigating this bias has become a central priority in recent research efforts (e.g., see \cite{Mayank_2023,Xiong_2025} and references therein). In our particular case, we will confront this problem directly, since the formation of black holes due to gravitational collapse leads to sharp gradients in the physical variables, becoming of utmost importance to address the issue particularly in scenarios near criticality and in supercritical regimes. In this context, Gaussian embeddings~\cite{Tancik_2020} project the physical input coordinates into a higher-dimensional space, thus alleviating the problem and facilitating the resolution of high frequencies. Other types of coordinate transformation, such as radial basis functions or Chebyshev polynomials~\cite{Huang_2025}, can also be considered. In our specific context, the embedding module constitutes a core element of the architecture, integrating multiple techniques designed to enhance the analytical capabilities of the model. It comprises two parallel and independent paths, which we describe in the following.

\begin{itemize}
    \item \textbf{Upper branch}. The input coordinates provided, $(t,r)$, will lie within the domain $[0,T]\times[0,R]$, where $T$ and $R$ denote the maximal temporal and spatial extents of the sampling, respectively, which are arbitrary depending on the problem at hand. Since arbitrarily large values are typically impractical for this class of neural networks, the initial step is to normalize the coordinates according to
    \begin{equation}
    \left(\tilde{t},\tilde{r}\right)=\left(\frac{t-t_{\mathrm{center}}}{t_{\mathrm{scale}}},\frac{r-r_{\mathrm{center}}}{r_{\mathrm{scale}}}\right)\in[-1,1].
    \label{eq:coords_norm}
    \end{equation}
    In this way, the coordinates are recast into a numerically well-conditioned interval for the PINN, defining the temporal center and scale as $t_{\mathrm{center}}=\left(t_{\mathrm{max}}+t_{\mathrm{min}}\right)/2=T/2$ and $t_{\mathrm{scale}}=\left(t_{\mathrm{max}}-t_{\mathrm{min}}\right)/2=T/2$; the spatial variable is treated analogously. Once normalized, the coordinates are passed through a ``warp'' block. Conceptually, this block applies a function of the form
    \begin{equation}
    g_{d}(\mathbf{x})=a_{d}\mathbf{x}_{d}+b_{d}+\sum_{k=1}^{K}c_{k,d}\:\mathrm{tanh}\left(s_{k,d}\left(\mathbf{x}_{d}-\nu_{k,d}\right)\right),
    \label{eq:tanh_warp}
    \end{equation}
    i.e.~an affine transformation, $ax+b$, augmented with a superposition of $\mathrm{K}$ independent bump-like deformations (scaled and shifted hyperbolic tangents). Because all parameters $(a,b,c_{k},s_{k},\nu_{k})$ are trainable, each bump introduces a smooth local distortion around its center $\nu_{k}$. By combining multiple bumps, the transformation can selectively stretch or compress different regions of the coordinate axes. Here, $d\in\{t,r\}$ refers to the coordinate dimension and $\mathbf{x}$ to the coordinate tensor.
    This transformation is particularly valuable for solving critical scenarios in which a black hole forms from gravitational collapse and both sharp and smooth features coexist within the same domain. By adapting the effective resolution of the coordinates, this block yields a smooth, infinitely differentiable, and fully adaptive representation in regions with sharp gradients, especially close to the origin. Beyond these transformations, the normalized coordinates are also used to generate some low-order polynomials such as $\tilde{t}^{2}$, $\tilde{r}^{2}$ or $\tilde{t}\:\tilde{r}$, which are concatenated to the previous features, thereby giving more information to the model, which may help overcome spectral bias~\cite{Shaghayegh_2025}.
    \item \textbf{Lower branch.} Simultaneously, before normalizing the coordinates, a dictionary of Gaussian radial basis functions $\{\varphi_{i}(t,r)\}_{i=1}^{M}$ is constructed. This dictionary comprises $\mathrm{M}$ different individual functions, each defined as
    \begin{equation}
    \varphi_{i}(\mathbf{x})=\mathrm{exp}\left[-\frac{1}{2}\sum_{d\in\{t,r\}}\left(\frac{\mathbf{x}_{d}-\mu_{i,d}}{l_{i,d}}\right)^{2}\right],\qquad i=1,\ldots,M,
    \label{eq:Gaussian_RBF}
    \end{equation}
    and centered at a learnable location $\mu_{i}=\left(\mu_{i,t},\mu_{i,r}\right)$ with learnable anisotropic scales $l_{i}>0$ for both time and space. Because all parameters are trainable, the model learns \textit{where} and \textit{how wide} to place these Gaussian characteristics according to the physical cost function. By stacking $\mathrm{M}$ of such functions, the embedding yields a nonlinear, high-dimensional, and smooth representation of the original input space. This enriched map captures localized structure and facilitates the detection of regions with sharp gradients, such as those taking place near formation of black holes.
\end{itemize}

Finally, the features produced by both branches are concatenated and projected through two consecutive quadratic layers to generate the output of the block.

\textbf{Quadratic residual layers.} Motivated by the promising results obtained by our numerical experiments, we propose using quadratic residual layers~\cite{qres_2021} in place of simple linear layers. These implement the straightforward operation
\begin{equation}
z=(\mathbf{W}_{2}x)\:\circ\:(\mathbf{W}_{1}x)+\mathbf{W}_{1}x+b,
\label{eq:QRes_layer}
\end{equation} prior to nonlinear activation, introducing two sets of trainable weights, $\mathbf{W}_{1}$ and $\mathbf{W}_{2}$. Therefore, for problems where linear layers suffice to capture the functional complexity of the target function, one would expect $\mathbf{W}_{2}\to 0$. Conversely, when the model takes advantage of the second-order components, these weights will be non-negligible. Here, ``$\circ$'' stands for the element-wise matrix multiplication (a.k.a. Hadamard product).

A central point behind the motivation is that many physical mappings are nonlinear even before the activation function. Equipping the model with a second-order interaction allows it to natively express numerous multiplicative couplings (e.g., products of variables and their derivatives) that may appear directly in the PDEs. In our particular equations, many terms are at least of second order in the variables---for example, products such as $\alpha\Phi$ or $\alpha\Pi$---and quadratic contributions in energy densities like $X^{2}+Y^{2}$. QRes pre-activations can naturally handle these structures, reducing the number of factorizations that linear layers would otherwise need to learn, typically improving general efficiency and mitigating residual stiffness. For resolving criticality, the extremely sharp gradients observed in the physical variables near the black hole threshold demand high model expressivity. These layers can be particularly advantageous in this regime, offering greater resolving power without a blow-up in the number of parameters.

\textbf{Adaptive activation functions.} Activation functions are essential in neural networks to endow models with nonlinearity, allowing them to address general problems rather than degenerate into simple successive linear combinations of the input. Depending on the chosen activation, these functions can also be used to constrain the physical ranges of the variables of interest. In our specific setting, the final set of physical variables comprises the field $\phi$, the lapse function $\alpha$, and the compactness parameter $C$. The field $\phi$ is not confined to any certain interval, so no activation is necessary for this output (the identity function is used). On the other hand, both $\alpha$ and $C$ are physically restricted to the range $[0,1]$; in particular, values of $C\to 1$ represent the emergence of a black hole singularity.

Beyond constraining physics to the admissible domain of the problem at hand, activation functions also play a fundamental methodological role. Their choice can be critical to the performance of DL models and, in particular, of PINNs when solving physical scenarios. The influence of activation functions on model performance has attracted considerable attention in the recent literature. Some studies~\cite{Maczuga_2023} analyze their effects on relatively simple problems, while others~\cite{Abbasi_2024} design problem-specific activations tailored to the physics; an approach that, while potentially effective, may introduce excessive dependence on the problem. A widely shared conclusion is that no single ``default'' activation function fits all the problems in PINNs, outperforming all others in different fields. Consequently, the choice of activation is highly problem-specific, as some authors~\cite{Honghui_Wang_2023} propose employing a set of basis activation functions, allowing the model itself to select the certain subset from which it can benefit the most.

In our setting, because the weight mappings of the hidden layers are quadratic, it is crucial to employ activation functions that act in a bounded range. Using unbounded activations such as SiLU or ReLU could introduce numerical instabilities by producing arbitrarily large values, particularly during the early stages of training. As noted before, a main concern in the field of PINN methodologies is mitigating their spectral bias, which often causes these models to be unable to resolve problems featuring rapid oscillations and sharp gradients, which is indeed something that we will encounter when solving gravitational collapses. Among the available strategies to counteract this effect, one is to provide the model with an additional degree of freedom by including trainable activation functions that the network can adapt to its needs (see, e.g.,~\cite{Afrah_2025} and references therein). These functions can be made trainable by introducing values that the network learns, with the specific parameterization depending on both the chosen function and the problem at hand. The parameters may be global, layer-specific, or even neuron-specific. In our case, the trainable activation for the hidden layers will be a hyperbolic tangent, 
\begin{equation}
\sigma_{\mathrm{hidden,i}}(x):=\mathrm{tanh}\left(\psi_{i}x+\xi_{i}\right),
\label{eq:trainable_tanh}
\end{equation}
where $\{\psi_{i},\xi_{i}\}$ are independent parameters for the $i$-th hidden layer: $\psi_{i}$ controls the slope while $\xi_{i}$ represents the offset. As noted, among $(\phi,\alpha,C)$ the field $\phi(t,r)$ is not constrained, but $\alpha$ and $C$ are physically restricted to the interval $[0,1]$. Therefore, the most natural choice for these two is a (trainable) sigmoid activation
\begin{equation}
\sigma_{\mathrm{output}}(x):=\mathrm{sigmoid}\left(\psi_{\mathrm{output}}x+\xi_{\mathrm{output}}\right).
\label{eq:trainable_sigmoid}
\end{equation}
The set $\{\psi_{\mathrm{output}},\xi_{\mathrm{output}}\}$ can be shared or distinct for the variables; in all our experiments, however, we will treat them as different. One might also consider a learnable scaling parameter that multiplies the entire function; however, this can be counterproductive, as it may incorrectly modify the admissible physical ranges of the variables.

For the gravitational collapse problems that we aim to tackle, activation functions---particularly those in the hidden layers---are not a central concern. However, making them trainable can provide the model with an additional degree of flexibility, even if they are not decisive. Relative to a vanilla PINN, the modification is modest: we introduce only two extra parameters per hidden layer and four at the output (two for each trainable head), resulting in a negligible computational overhead. That said, care is needed: an excessive number of additional parameters can induce numerical instabilities and substantially slow training without yielding better solutions.

\section{Results}
\label{sec:Results}

In this section, we propose several initial-value problems for the scalar field $\phi(t,r)$, each leading to an independent solution and encompassing both the subcritical and supercritical regimes of gravitational collapse. We solve these problems using a range of recent methodologies from the literature---the ModPINN proposed in the previous section is included---to enable a systematic comparison across different approaches. In summary, this section is organized as follows: First, we introduce and discuss the problems to be solved and the metric used. Second, we outline the general parameter settings used for both the ModPINN and the comparison models, i.e., the setup of the methodologies and the technical details of the training process. Third, we report performance across all models and cases. Fourth, we present learning curves for the principal models. Fifth, we assess the robustness of our model with respect to the architectural choices and sample size. Sixth, we conduct a focused study of the critical regime and compare the methods performing the best. Finally, we visualize the resulting physical variables over the sampled spacetime.

\textbf{Problem setup.} As noted in the introduction, the seminal findings and conclusions of~\cite{Choptuik_seminal} are remarkable for their generality and universality. The study examines several families of initial conditions for the scalar field and concludes, for instance, that the exponent $\gamma$ in the black hole mass law~(\ref{eq:BH_mass_relation_3}) 
appears to be a universal quantity, independent of the initial field configuration. Instead, its dependence is directly related to the deviation of a control parameter $p$ from its critical value $p^{*}$. Despite the well-established universality of critical phenomena in gravitational collapse, in our experiments we fix the initial field profile to the ``family (a)'' (a.k.a. $\phi_{0}(r)$ from now on) configuration from~\cite{Choptuik_seminal}, expressed as 
\begin{equation}
\phi(0,r)=\phi_{0}r^{3}\:\mathrm{exp}\left(-\frac{(r-r_{0})^{q}}{\delta^{q}}\right).
\label{eq:initial_phi_shape}
\end{equation}
This field profile corresponds to a compact Gaussian pulse for which the scaling law involving the mass of the black hole, $M_{\mathrm{BH}}$, is known to persist well beyond the asymptotic limit $p\to p^{*}$~\cite{Choptuik_1998}. Specifically, the profile is fixed by the set of parameters $\{\phi_{0},r_{0},\delta,q\}$. Among these, the parameter of interest that we vary towards criticality is the initial amplitude $\phi_{0}$; from now on, ``$p$'' will be identified with $\phi_{0}$, and the critical value ``$p^{*}$'' identified as the critical amplitude. The remaining parameters are fixed for all tests: $r_{0}=0$, $q=2$, and $\delta=\sqrt{2}\sigma$ with $\sigma=1$.

This profile constitutes the only prior information that is available before any numerical solution---whether via DL or FDMs---is obtained. However, the initial profiles for the other variables, $\alpha(t,r)$ and $C(t,r)$, are also required for the PINN to learn. These are calculated by setting $t=0$ in the field equations~(\ref{eq:constraint_1}--\ref{eq:constraint_3}) and solving numerically as a preprocessing step. It should be noted that, once the initial pulse is specified, the spatial derivative of the field, $\frac{\partial\phi}{\partial r}$, is straightforwardly determined, while its temporal derivative is not. Because the field evolution equations~(\ref{eq:constraint_1}) are second order, the latter must be imposed. Several choices are possible; for example, $\frac{\partial\phi}{\partial t}=0$ (time-symmetric initial data, $\Pi=0$), or $\Pi=\Phi$ (purely ingoing field). In this work, we adopt $\Pi=0$ for all problems and numerical solutions considered. This choice is purely pragmatic: for mathematical simplicity---both in the DL model and in generating the numerical solution---we set $\Pi=0$ without entailing any loss of generality for the study. Regarding the boundary conditions, we impose $C(t,0)=0$ (the so-called elementary flatness at the origin). For $\alpha(t,0)$, there is gauge freedom: $\alpha(t,0)$ may be chosen arbitrarily and $\alpha(t,r)$ can be rescaled by a constant, so that at $r=R$ one may enforce $\alpha(t,R)=\frac{1}{\alpha(t,R)}$. In our setting, we adopt derivative-based boundary conditions~(\ref{eq:falloff_conditions}) instead of Dirichlet conditions, since fixing these variables to constant values might introduce problem-specific dependence. Moreover, since our main objective is to resolve criticality in gravitational collapse using DL-based methods, the chosen scenario is sufficient: it exhibits the relevant interesting behavior and we have a reasonably accurate estimate of the critical parameter, enabling us to distinguish subcritical from supercritical regimes. This setup therefore supports a comprehensive study of the problem, despite the fact that the space of admissible initial field profiles is theoretically unlimited.

Regarding the evaluation metric, we employ the relative mean squared error, 
\begin{equation}
l_{2}=\mathop{\scalebox{2.5}{$\sum$}}\limits_{i\:\in\:\{\phi,\alpha,C\}}\frac{\sqrt{\displaystyle\sum_{(t,r)\:\in\:[0,T]\times\Omega}\left|\pmb{\mathcal{F}}^{i}_{\Theta}(t,r)-\pmb{\mathcal{F}}^{i}_{\mathrm{GT}}(t,r)\right|^{2}}}{\sqrt{\displaystyle\sum_{(t,r)\:\in\:[0,T]\times\Omega}\left|\pmb{\mathcal{F}}^{i}_{\mathrm{GT}}(t,r)\right|^{2}}},
\label{eq:l2_definition}
\end{equation}
between the prediction of the model, denoted as $\pmb{\mathcal{F}}_{\Theta}$, and the ground truth solution, $\pmb{\mathcal{F}}_{\mathrm{GT}}$. In particular, this metric is simply the sum of the relative errors $l_{2}$ for each output variable, so one may also define variable-wise errors $l_{2}^{\phi}$, $l_{2}^{\alpha}$, $l_{2}^{C}$, which can be useful for dissecting performance.

\textbf{General configuration.} As outlined in the methodology, numerous parameters and design choices can be specified regarding the configuration of our PINN model (ModPINN). Therefore, it is necessary to define a ``general configuration'', which will be used in all the cases studied unless otherwise stated.

\begin{itemize}
    \item \textbf{Loss function.} Regarding the loss computation, we set the weights in~(\ref{eq:Loss_vs_t_r}) as follows: $\lambda_{\mathcal{IC}}=10^{3}$ for the initial conditions; $\lambda_{\mathcal{BC},\mathrm{left}}=\lambda_{\mathcal{BC},\mathrm{right}}=1$ for the boundary conditions at $r=0$ and $r=R$; and $\lambda_{i}=1$ for all the field equations, that is, the PDE residuals. Although several studies conclude that dynamically adapting these weights to the loss of each term can be beneficial, we opt to avoid unnecessary additional complexity and keep them fixed, with $\lambda_{\mathcal{IC}}\gg\{\lambda_{i}\}$ to ensure accurate learning of the initial conditions, which largely determine the physical solution. Regarding the causality and spatial enforcement terms, we use $(\epsilon_{\mathrm{t}},\epsilon_{\mathrm{r}})=(50,10)$. These values are neither excessively large nor particularly small, leading to a necessary, though nontrivial modification of the vanilla loss function.
    \item \textbf{Spacetime sampling.} With respect to the sampled spacetime domain $(t,r)$, several configurations may be considered. In general, the number of points does not need to be large; a standard grid of $N_{\mathrm{t}}\times N_{\mathrm{r}}=128\times 128$ is typically sufficient. Lighter $(64\times 64)$ and heavier $(256\times 256)$ grids are also examined, particularly to assess how the sampling density affects the model performance. For adaptivity and point movement, we adopt a modest value $\Lambda=1$ in~(\ref{eq:m_density}), which allows light repositioning of the points.
    \item \textbf{Neural architecture.} Regarding the network layout, we employ quadratic-weight layers, stacking six hidden layers with 50 neurons each after embedding. For the embedding itself, the upper branch uses $K=32$ hyperbolic-tangent bumps for each input coordinate, while the lower branch includes a dictionary of $M=32$ Gaussian radial basis functions. These choices allow us to exceed the simplicity of vanilla PINNs without imposing substantial additional cost. Finally, after concatenating both branches, we use a 50-neuron hidden layer to project the combined features to the output of the block. Regarding the random initialization of trainable parameters in the linear layers, we adopt the default setting in PyTorch, namely the Kaiming He uniform initialization scheme~\cite{Kaiming_2015}. For the function $g_{d}(\mathbf{x})$, the parameters $a_{d}$ and $b_{d}$, which are shared across all bumps, are initially set to 1 and 0, respectively. The $\{s_{k,d}\}$ are initialized to 1 or all $k$, while the $\{\nu_{k,d}\}$ are uniformly initialized over the interval $[-1,1]$ partitioned into $k$ segments. Similarly, for the lower branch, the set $\{\mu_{i,d}\}$ is sampled randomly from a uniform distribution over $[-1,1]$, while all elements of $\{l_{i,d}\}$ are initially assigned to 1. Finally, for the trainable activation functions in~(\ref{eq:trainable_sigmoid}) and~(\ref{eq:trainable_tanh}), the parameter sets $\{\psi_{i}\}$, $\psi_{\mathrm{output}}$ and $\{\xi_{i}\}$, $\xi_{\mathrm{output}}$ are set to 1 and 0, respectively, for all hidden layers and for both physical outputs.
    \item \textbf{Training details.} The optimizer employed in all training runs and experiments has been Shampoo with Adam in the Preconditioner's eigenbases (SOAP)~\cite{SOAP_2025}, a recently introduced method in the optimization literature. SOAP has achieved state-of-the-art results in general tasks such as language modeling. More interesting to our work, recent studies~\cite{Sifan_Wang_2025} have applied SOAP to solve various PDE problems with PINN-type models, reporting highly competitive performance and even surpassing commonly used baselines such as Adam or AdamW. Nevertheless, this optimizer is effectively of (quasi) second-order, which typically means slightly slower training per iteration; however, this is countered by faster and more accurate convergence. Unless otherwise specified, we train for 100,000 epochs with a fixed learning rate of $10^{-4}$ and normalize the gradient norm to unity. If overfitting is detected---manifested as degradation in the physics loss---we retain the best-performing checkpoint for each case, the selected models being the ones reported in the analyses.
\end{itemize}

\textbf{Numerical comparisons.} The field equations~(\ref{eq:constraint_1}-\ref{eq:constraint_3}) are straightforwardly solvable in subcritical regimes, i.e., in the absence of black holes, where a vanilla PINN is typically enough. However, as the control parameter $p$ (here, the initial amplitude $\phi_{0}$) approaches its critical value $p^{*}$, the solution develops increasingly sharp and localized gradients (in both space and time). This regime demands a more exhaustive analysis and a modeling specifically tailored to handle such extreme behaviors. Given the inherent difficulty of modeling gravitational collapses, we refrain from relying on a single model. Instead, we compare solutions produced by several PINN-based approaches proposed in recent literature, encompassing methodologies such as the KANs~\cite{kan_2025}, the PirateNets~\cite{piratenets_2024} or the PINN ``FP64''~\cite{chenhui_2025} which refers to a vanilla PINN where double-precision floating-point format has been considered. Other methodologies---such as PINNsFormer~\cite{pinnsformer_2024}, which is based on attention, and the recent PINNMamba~\cite{mamba_2025}, which introduces subsequence modeling via State Space Models (SSMs)---were also considered but excluded from the final study due to their training and computational requirements being substantially higher while yielding performances similar to, or even worse than, other simpler models. All training was performed in PyTorch on an NVIDIA A100 (40GB) GPU.

The final results for all models considered and four distinct initial-condition cases are reported in Table~\ref{tab:table_l2}. From previous numerical studies~\cite{Choptuik_seminal,Choptuik_1998}, we know with high accuracy that the critical value of the parameter $\phi_{0}$ in the proposed problem occurs near $\phi_{0}=0.1125$, which we adopt as the critical scenario of gravitational collapse. Consequently, we examine two subcritical settings ($\phi_{0}=0.1,0.10625$) and one supercritical setting ($\phi_{0}=0.125$). The exact same loss function, with parameters $(\epsilon_{\mathrm{t}},\epsilon_{\mathrm{r}})=(50,10)$, is used for all models, as is the adaptive spacetime sampling prescribed by our methodology with $\Lambda=1$. Therefore, the purpose has been to compare neural architectures to handle the gravitational collapse while keeping all other factors fixed. Regarding model-specific settings, we note that, where feasible, all architectures employ six hidden layers with 50 neurons each---except PirateNet, which is organized in blocks rather than layers, and KAN, which uses three layers with five nodes each. Where possible, the hyperbolic tangent is used as an activation function for the hidden layers. Finally, for the proposed ModPINN, the previously introduced standard configuration is adopted.

\begin{table}[!htbp]
  \centering
  \scriptsize
  \caption{Best achieved values of the relative total $l_{2}$ error~(\ref{eq:l2_definition}) after 100,000 epochs across state-of-the-art models for the scalar-field initial-value problems. The best results for each case are shown in bold.}
  \label{tab:table_l2}
  \begin{tabular}{l|cccc}
    \toprule
    \textbf{Model} 
      & \multicolumn{4}{c}{$\phi_{0}$} \\
    \cmidrule(l){2-5}
      & 0.1    & 0.10625 & 0.1125  & 0.125   \\
    \midrule
    \textbf{PINN} & $0.119\pm0.009$ & $0.160\pm0.013$ & $0.694\pm0.210$ & $0.709\pm0.340$ \\
    \textbf{PINN ``FP64''~\cite{chenhui_2025}}              & $0.114\pm0.005$ & $0.165\pm0.010$ & $0.710\pm0.240$ & $0.863\pm0.610$ \\
    \textbf{sf-PINN}~\cite{Wong_2024}        & $0.113\pm0.006$ & $0.138\pm0.017$ & $0.960\pm0.290$ & $0.855\pm0.320$ \\
    \textbf{QRes}~\cite{qres_2021}     & $\pmb{0.111\pm0.009}$ & $\pmb{0.132\pm0.008}$ & $0.817\pm0.203$ & $0.333\pm0.160$ \\
    \textbf{KAN}~\cite{kan_2025}       & $0.456\pm0.250$ & $0.498\pm0.320$ & $1.888\pm0.190$ & $2.107\pm0.210$ \\
    \textbf{PirateNet}~\cite{piratenets_2024} & $0.133\pm0.009$ & $0.151\pm0.030$ & $0.698\pm0.240$ & $0.566\pm0.180$ \\
    \textbf{ModPINN}              & $0.114\pm0.006$ & $0.152\pm0.015$ & $\pmb{0.588\pm0.060}$ & $\pmb{0.297\pm0.090}$ \\
    \bottomrule
  \end{tabular}

\end{table}

For each case and each model, we have conducted 10 independent training runs, randomizing the seed---and thus the initial parameters---in every instance, thereby generating some statistics. We report the mean performance alongside the standard deviation. Inspecting the results, we observe that the critical and supercritical cases exhibit the largest variability across most models, highlighting the difficulty in solving these scenarios. For the critical case, ModPINN achieves the best performance in both mean accuracy and variability, indicating greater stability than the alternatives for this scenario. The same conclusion holds for the supercritical case with $\phi_{0}=0.125$. In contrast, in the two subcritical settings, the purely quadratic PINN (QRes) leads to the strongest results; however, as noted, these cases are easily solved by a vanilla PINN due to their relative simplicity and smoothness, so the outcome and the similarity across models are generally accepted. It is therefore expected that QRes ranks among the strongest models: quadratic-weight layers help mitigate the frequency bias typical of PINNs. These layers also aid in resolving sharp localized gradients, as evidenced by the proposed ModPINN. However, for $\phi_{0}=0.1125$ and $0.125$, QRes is not among the top performers. The principal reason is that, while quadratic layers can substantially benefit such problems, they may also induce numerical instabilities. Therefore, implementing the proposed embedding---which normalizes input coordinates to numerically manageable values and generates additional features---indeed improves performance on criticality and gravitational collapse cases. These are precisely the regimes of greatest interest as they reveal the major physical phenomena.

For the remaining models, performance in subcritical cases is generally consistent across architectures, with the exception of the KAN considered, which is slightly behind the others. This behavior may be a consequence of its chosen architectural hyperparameters, as we tried to find a balance between computational cost and accuracy, since it is known that KANs tend to train more slowly. It should be noted that the results reported here reflect a single configuration of the methodology beyond architecture alone. With a careful study of hyperparameter tuning, better outcomes could likely be achieved for all methodologies, particularly those that perform poorly in this comparison.

\begin{figure*}[t]
    \centering
    \includegraphics[width=1.75\columnwidth]{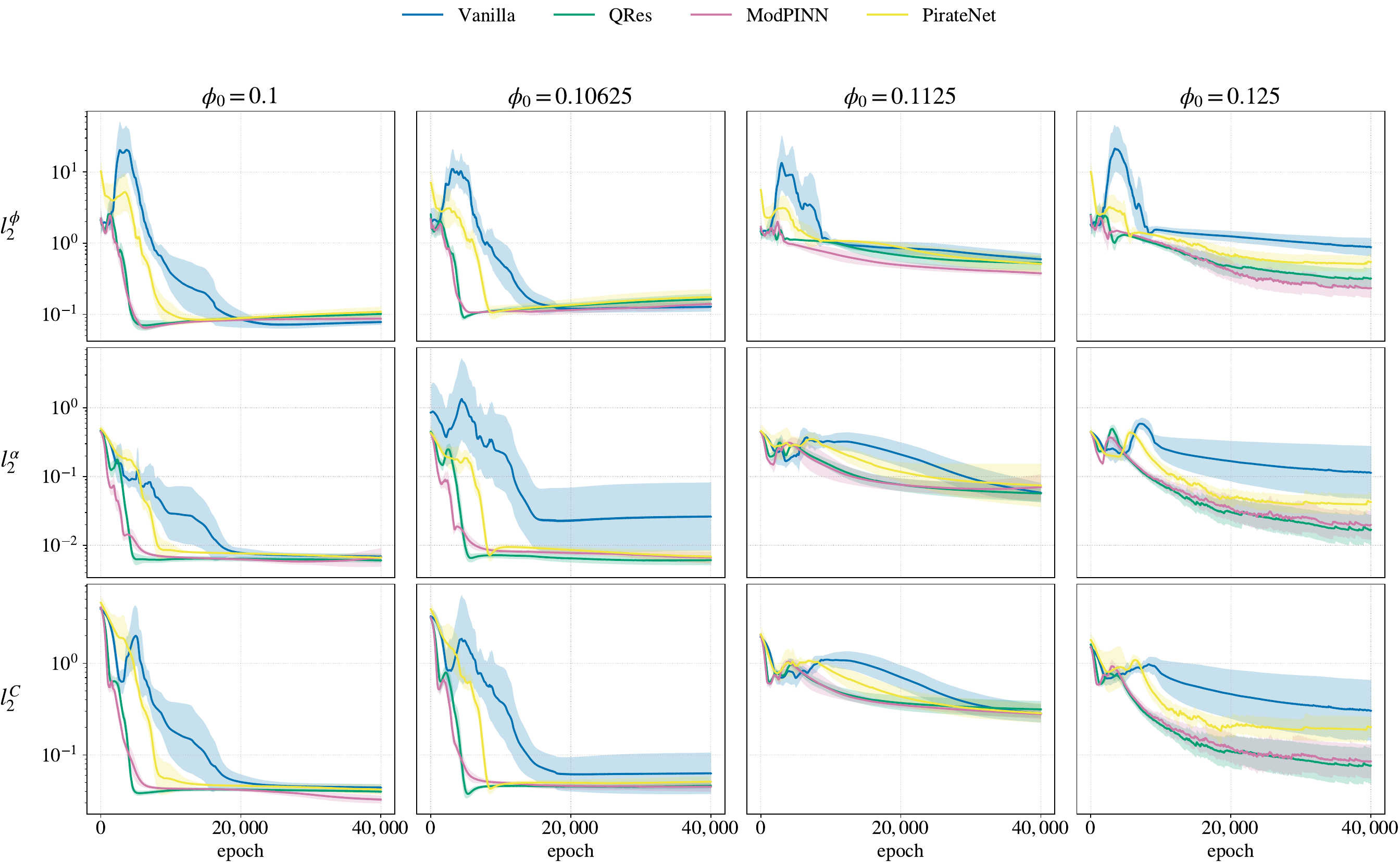}
    \caption{Convergence of the relative $l_{2}$ errors for each physical variable (row by row) over training, across different models and initial-value problems. Beyond approximately $40,000$ epochs, performance gains become barely minimal.}
    \label{fig:results_3}
\end{figure*}

\textbf{Training curves.} As noted earlier, the relative quadratic errors $l_{2}$ defined in~(\ref{eq:l2_definition}) can be decomposed into a sum of components, $l_{2}=l_{2}^{\phi}+l_{2}^{\alpha}+l_{2}^{C}$. This decomposition allows us to assess the performance of each variable individually, rather than relying solely on the total sum, which may lead to masked deficiencies in training. In figure~\ref{fig:results_3}, we present the learning curves for these metrics across all considered problems, focusing on the four best performing models. For each variable, the error scale is kept fixed in all different scenarios. The first conclusion is that the cases showing gravitational collapse, especially the critical case ($\phi_{0}=0.1125$), are by far the most challenging to train, achieving higher errors in general and evolving within a much narrower range. In general, all models achieve comparable performance in this case. It is worth highlighting that, given the construction of the $l_{2}$ metric, small variations can translate into substantial improvements in the target variable---particularly in regions exhibiting sharp changes over small $(t,r)$ intervals. Moreover, even though most models converge faster than the vanilla PINN, none stands out as decisively superior---except ModPINN, which yields an average total error of 0.588, compared with 0.694 and 0.698 for the next best (the vanilla and PirateNet, respectively) and with variability roughly three times smaller. In principle, given the difficulty of the critical case, likely due to a sharp and hard-to-navigate loss landscape, we might expect independent trainings to yield slightly different results. However, as Table~\ref{tab:table_l2} has shown, the variability is concerning across models, on the order of $\sim30\%$ relative to the mean, except for ModPINN, where it is only $\sim10\%$. This indicates a more stable model that may admit further optimization and potentially even better results with a more exhaustive preconditioning of the hyperparameters. For the supercritical case, $\phi_{0}=0.125$, it seems that all models identify black hole formation more clearly as the dominant result, as evidenced by the errors being uniformly lower than those of the vanilla PINN. In addition, the learning curves suggest that the problem becomes less difficult with a substantially wider range of progress during training. Finally, for the subcritical cases, all models give similar solutions, except for $\alpha(t,r)$ produced by the vanilla PINN at $\phi_{0}=0.10625$. We also observe that every alternative model converges faster---roughly in half the training time---representing a clear advantage. However, in general, the main conclusion is that all methods perform reasonably well due to the relative ``ease'' of these problems: smooth gradients make them straightforward for any of the models to solve.

It is important to note that these learning curves correspond to the training runs reported in table~\ref{tab:table_l2}; thus, all methodological settings are the same, including the spatiotemporal sampling size $N_{\mathrm{t}}\times N_{\mathrm{r}}=128\times 128$. Several factors could be adjusted to improve results: increasing sampling density, fine-tuning $\epsilon_{\mathrm{t}}$ and $\epsilon_{\mathrm{r}}$, and / or refining the weights assigned to the initial and boundary conditions. For the critical case in particular, a more detailed analysis is needed, which we will present in the remainder of the paper.

\textbf{Robustness of ModPINN.} Up to this point, we have reviewed a set of recent models in the literature, but it is worth putting more focus on our own proposal, which we have denoted as ModPINN. In essence, this architecture is no more than a modified PINN that replaces linear layers with quadratic ones, and whose main strength is an embedding introduced before the layer concatenation. As we have shown, this embedding leads to improved results for the supercritical case and, specifically, for the critical case $\phi_{0}=0.1125$, improving average performance and reducing variability. Taking into account that this model is the one giving the strongest results for gravitational collapse, it is necessary to carry out a more thorough examination of how various pre-training design choices influence its performance. Among these factors, we include the sampling size (that is, the number of spatiotemporal sampling points, $N_{\mathrm{t}}\times N_{\mathrm{r}}$) and the scale of the neural architecture, in particular the number of quadratic layers and the number of neurons per layer. Another parameter of interest that we examine here is the strength of the temporal causality, $\epsilon_{\mathrm{t}}$, while $\epsilon_{\mathrm{r}}$ will be constant and fixed at 10 for this analysis.

\begin{figure*}[!t]
    \centering
    \includegraphics[width=0.9\textwidth]{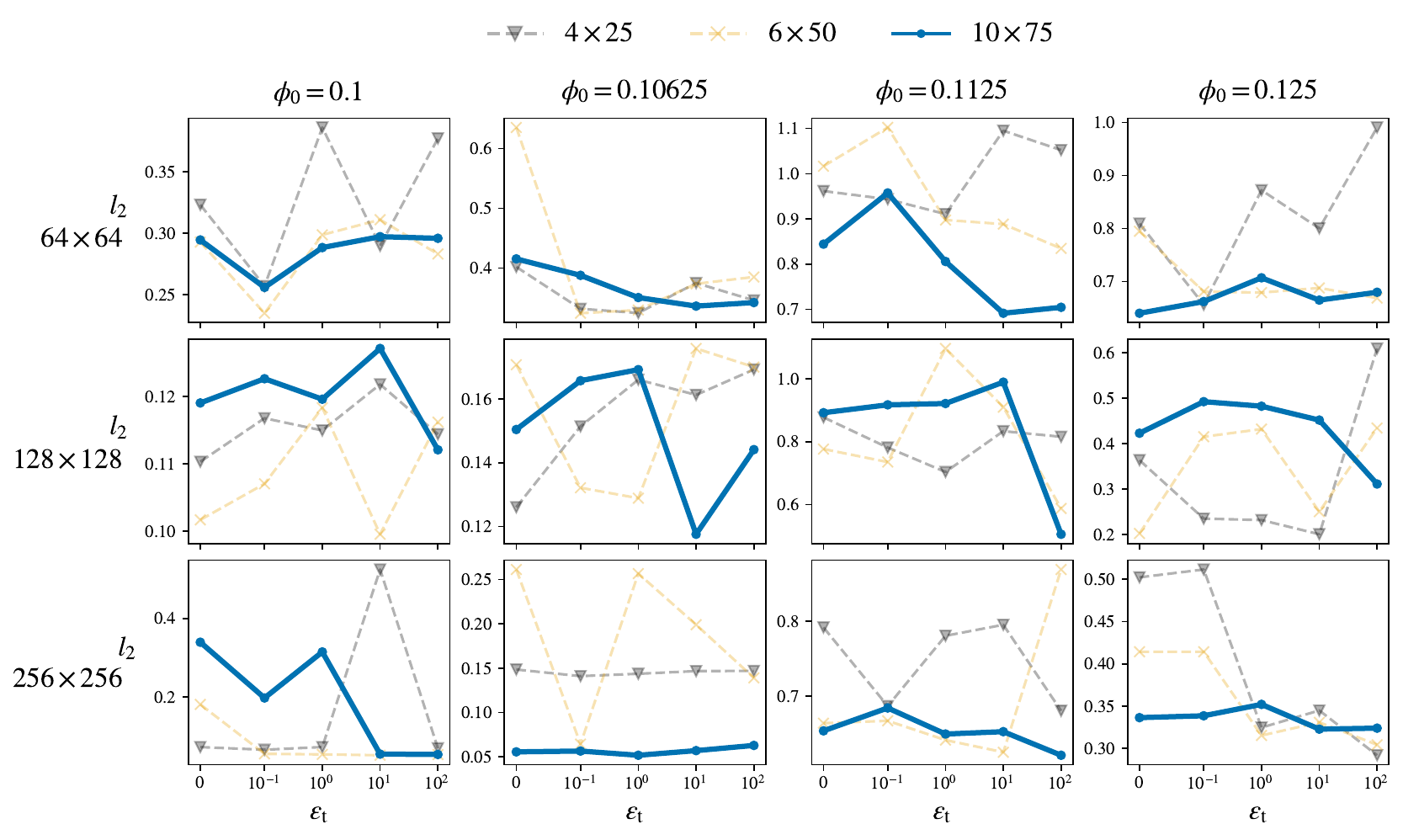}
    \caption{Relative total $l_{2}$ error of the proposed ModPINN for the considered scalar-field initial-value problems (by column) as a function of the hyperparameter $\epsilon_{\mathrm{t}}$, which modulates the strength of the temporal causality. Each row corresponds to a different spatiotemporal grid resolution $(t,r)$, and each sub-figure reports results for three network sizes.}
    \label{fig:results_1}
\end{figure*}

Figure~\ref{fig:results_1} reports the achieved $l_{2}$ as a function of $\epsilon_{\mathrm{t}}\in\{0,0.1,1,10,100\}$. Each row of sub-figures corresponds to one sampling size---$64\times 64$, $128\times 128$ and $256\times 256$---and within each sub-figure we display the errors for three different network sizes. Each column therefore corresponds to each initial-condition problem. By varying $\epsilon_{\mathrm{t}}$, we move the spectrum from a ``vanilla'' loss with respect to temporal causality to one that strongly enforces it at $\epsilon_{\mathrm{t}}=100$. Regarding sampling sizes, we consider grids from $64\times 64=4,096$ to $256\times 256=65,536$. Although the latter may appear large, we note that the original finite-difference simulations~\cite{Choptuik_seminal} employed on the order of $500,000$ grid points. In principle, then, the sampling size could be increased substantially; however, doing so would likely require batching strategies to feed the network, probably leading to higher computation times that are out of the main scope of this study. Several key insights can be concluded from this figure.

\begin{itemize}
    \item The $10\times 75$ architecture---the deepest and widest configuration---yields the most stable overall results across all cases, much like the largest sampling size, $256\times 256$, which also exhibits the greatest stability. Consequently, the model clearly benefits from larger sampling grids and higher network capacities, providing robustness independent of the problem considered. In contrast, smaller networks display greater variability as $\epsilon_{\mathrm{t}}$ changes. This behavior is a clear indication of the inherent physical complexity of the problem.
    \item In general, the motivation behind the enforcement of temporal causality is to prevent PINNs from ``looking ahead'' in time, that is, from resolving late-time points before early-time ones~\cite{Wang_2024}. This approach can be helpful in problems with a strong dependence on initial conditions, where the evolution of the system is fully determined by them. Consequently, it is reasonable to expect that in a smooth, easy-to-solve case like $\phi_{0}=0.1$, causality enforcement will have limited impact, which is essentially what we observe. For this problem, there is no clear trend in $l_{2}$ as $\epsilon_{\mathrm{t}}$ increases, because the solutions with $\epsilon_{\mathrm{t}}=0$ are generally good and any observed improvements are not substantial.
    \item Following the same logic, for the critical case $\phi_{0}=0.1125$, we observe a stronger dependence on $\epsilon_{\mathrm{t}}$, although the errors are generally higher due to the steep highly localized gradients mentioned. We see that the model clearly benefits from enforcing causality, particularly for the $128\times 128$ case, where the error quickly drops for $\epsilon_{\mathrm{t}}=10^{2}$. Regarding the $\phi_{0}=0.125$ scenario---specifically with $256\times 256$ points and a $10\times 75$ network---the solution is markedly stable, showing only minor sensitivity to $\epsilon_{\mathrm{t}}$. By contrast, when smaller networks are used, the improvement becomes appreciable.
\end{itemize}

These results demonstrate that the model benefits from large sampling grids as well as deep and wide network architectures. While adaptive sampling can indeed be helpful when the number of points is relatively small, in general a denser spacetime regime helps the model to solve the task better. Moreover, enforcing causality generally improves stability, and it is especially beneficial in cases where strongly tuning the initial conditions is critical, as demonstrated for $\phi_{0}=0.1125$.

\begin{figure*}[!t]
    \centering
    \includegraphics[width=0.8\textwidth]{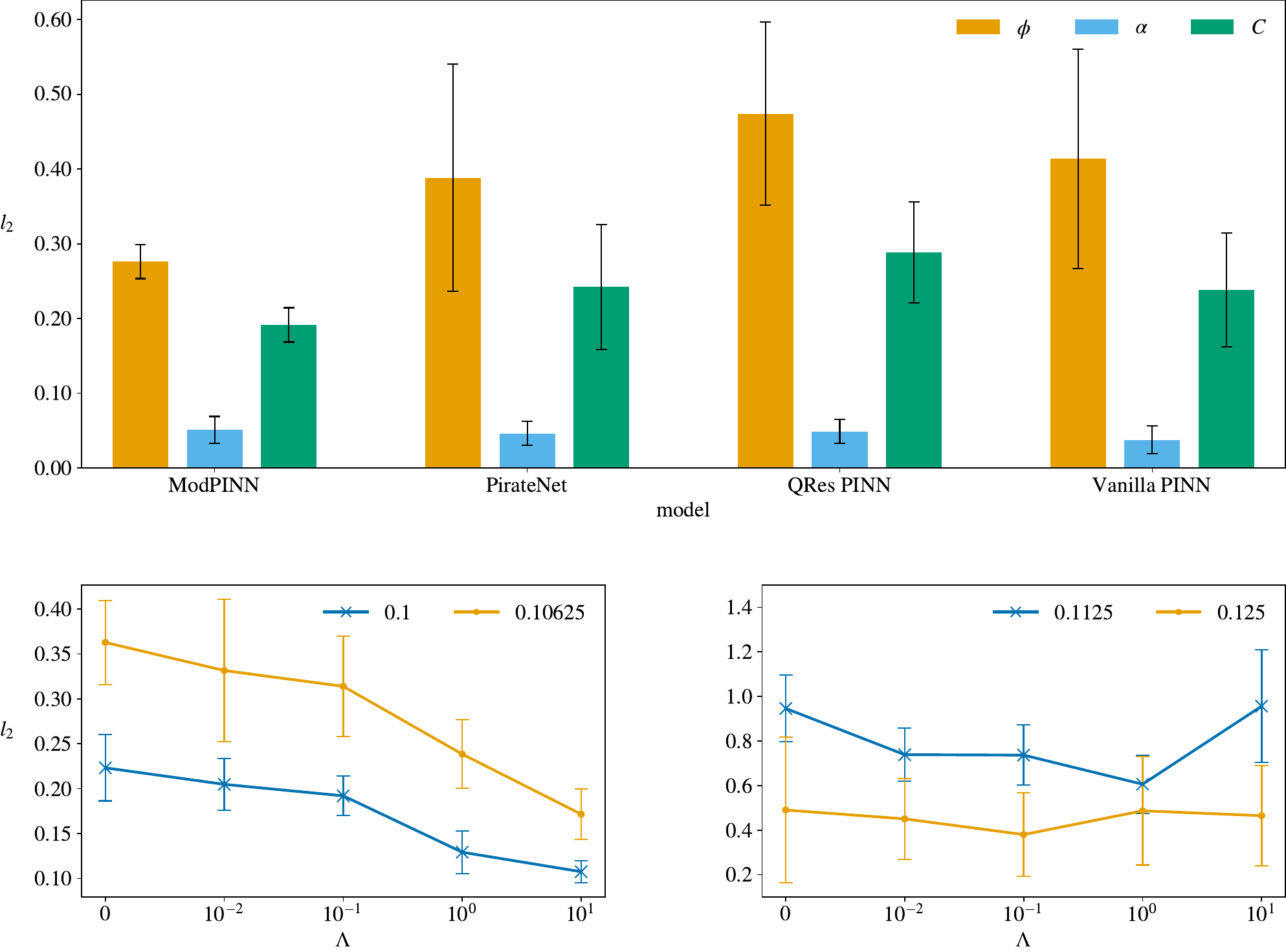}
    \caption{\textit{Top}: $l_{2}$ errors for each variable, together with their variability across ten independent runs, shown for different models considered. \textit{Bottom}: total $l_{2}$ errors as a function of the adaptive-sampling strength $\Lambda$, evaluated for each initial-condition problem. Only the ModPINN model is considered.}
    \label{fig:results_4}
\end{figure*}

\textbf{Critical regime.} At this point, it is important to focus on the analysis of the critical case for this family of initial data. This scenario, defined by $\phi_{0}=0.1125$, represents the boundary between dispersive fields and black hole formation. Accordingly, accurately resolving this regime is of particular interest for the community, paying special attention to the models that have yielded the most promising results: the QRes model, the PirateNet, the surprisingly effective vanilla PINN, and, finally, the ModPINN proposed in this review.

In figure~\ref{fig:results_4} (top), we show the mean relative $l^{2}$ error of each physical variable, together with its variability between several random trains, for all models considered in the critical case. In general, the lapse function $\alpha(t,r)$ is the best resolved quantity, having modest variability and slightly better performance in the vanilla PINN. This behavior can be attributed to its generally smooth profile, which does not present sharp gradients, making it easier to be approximated by any model. Regarding the other two variables, ModPINN achieves better performance than other state-of-the-art proposals, with our main claim focused on model stability. Across multiple independent trainings with random initializations, we observe that the convergence of our model is substantially more robust. Near the threshold of collapse, it is natural to expect a sharp loss landscape, such that small differences during optimization can lead to relatively different outcomes. However, ModPINN successfully attenuates this variability by yielding a more deterministic methodology for these problems in particular.

In the two bottom panels, we depict how the $l^{2}$ errors vary with the hyperparameter $\Lambda$, which controls the ``strength'' with which spacetime points $(t,r)$ are moved towards regions exhibiting higher values of the normalized physical loss, as defined in~(\ref{eq:m_density}). The lower-left panel reports this magnitude for the two subcritical cases ($\phi_{0}=0.1,0.10625$), while the lower-right panel presents the corresponding results for the critical and supercritical cases ($\phi_{0}=0.1125,0.125$). We observe that as $\Lambda$ increases within the range considered, the error shows a clear decreasing trend for the subcritical scenarios, with a particular reduction in variability at $\Lambda=10$ for the case $\phi_{0}=0.1$. For the critical scenario $\phi_{0}=0.1125$, the trend is also downward from $\Lambda=0$ to 1, with the error dropping from approximately $\sim 1.0$ to $\sim 0.6$. This constitutes a substantial improvement that should not be eclipsed by the use of a broader scale relative to the other panel. Regarding $\Lambda=10$ we observe a sudden increase in error for this case, indicating that the solutions deteriorate when this parameter is overly enlarged. This behavior is somewhat expected: the physics lies near the boundary of criticality, and the collocation points are required and used within a highly localized region of spacetime mainly determined by the black hole formation. Excessive and rapid relocation of the points can move some of them into undesirable regions, thereby devaluing their utility and, overall, leading to worse solutions. Finally, for the supercritical case with $\phi_{0}=0.125$, the trend is essentially flat. It is worth noting that the model under consideration here is ModPINN, which---consistent with table~\ref{tab:table_l2}---provides the best performance for this regime with quite low variability. Consequently, this scenario is already well resolved without additional techniques, as the model alone effectively minimizes the loss. It is therefore unsurprising that the solution does not improve substantially when adaptive sampling is introduced. By contrast, performing the same analysis with models such as the vanilla PINN---where this case is less accurately captured---might reveal a decreasing trend under the said sampling methodology.

\begin{figure*}[t]
    \centering
    \includegraphics[width=0.75\textwidth]{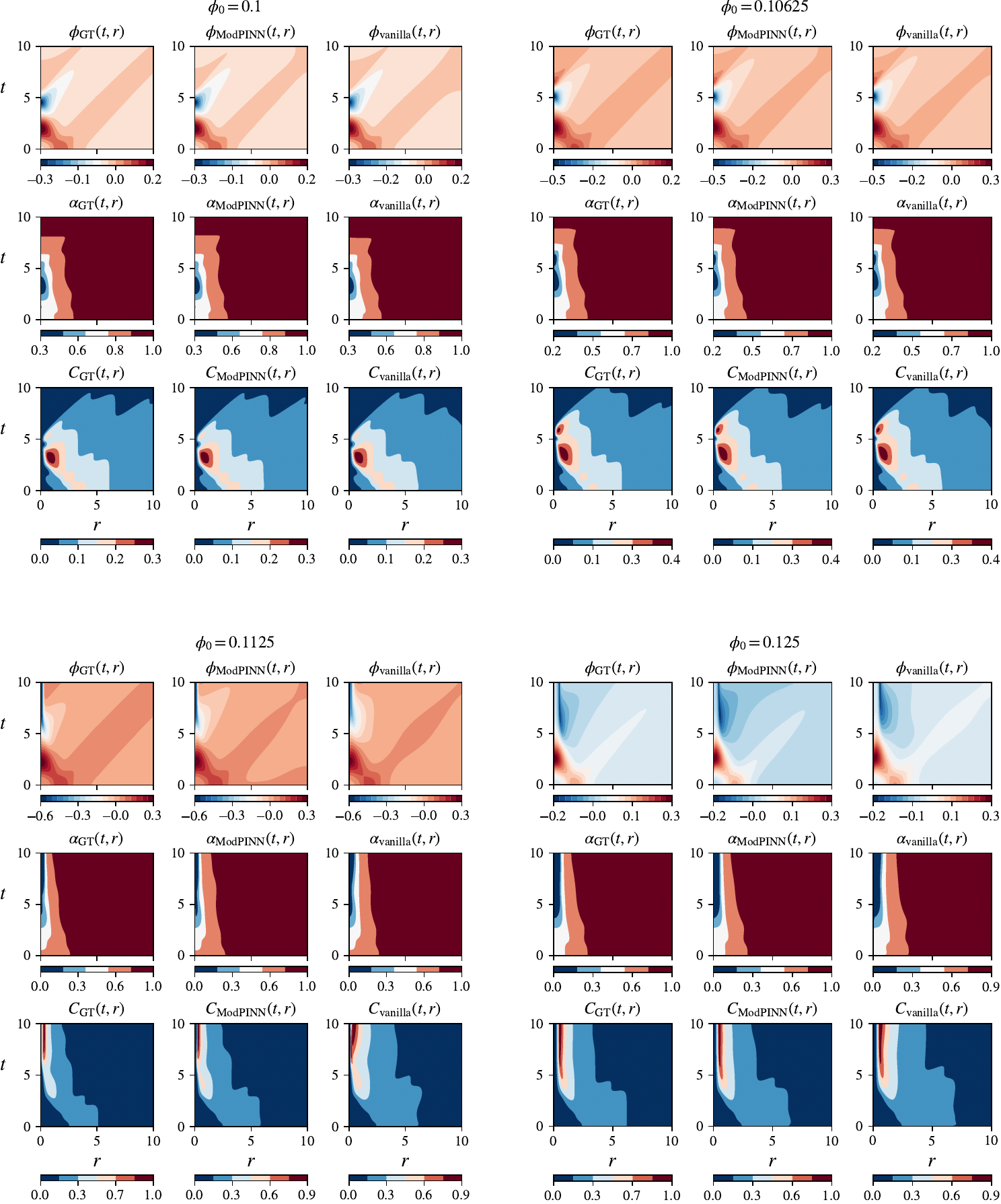}
    \caption{Two-dimensional visualizations of the physical variables $(\phi,\alpha,C)$ depicted over the sampled spacetime for each initial-condition case, $\phi_{0}\in\{0.1,0.10625,0.1125,0.125\}$. For each problem, the columns---from left to right---display the ``ground truth'' solution obtained via finite differences, the solution produced by the ModPINN, and the solution from a vanilla PINN, respectively. All cases employ $128\times 128$ sampling points and a standard configuration for both models.}
    \label{fig:results_2}
\end{figure*}

\textbf{Evolution of the physical variables.} Figure~\ref{fig:results_2} displays the physical variables---the scalar field $\phi$, the lapse function $\alpha$, and the compactness parameter $C$---over the spacetime domain $(t,r)\in[0,10]\times[0,10]$. All four initial-condition scenarios are shown. By columns, we present the ground truth solution, followed by the predictions from ModPINN and from the vanilla PINN, respectively. Perhaps the most interesting feature is the parameter $C$, which approaches 1.0 only in regions where an apparent horizon arises. As seen for $\phi_{0}=0.1125$ and $0.125$, the coordinate singularity emerges at times $t\gtrapprox 5.0$ with $r$ values near the origin. As $\phi_{0}$ increases beyond 0.1125, the black hole seems to shift to $r>0$. In both scenarios, the region where $C\sim 1$ is narrowly localized in $r$, a structure that ModPINN reproduces notably well. Although the vanilla PINN can detect this feature as well, it moves the coordinate singularity towards smaller $r$, yielding an overall inferior result. However, when using only the loss function~(\ref{eq:Loss_vs_t_r}) together with the adaptive sampling, the vanilla PINN is still able to identify the black hole with relatively strong performance. Regarding the other variables, we find that they are generally well resolved across all cases and for both models. The expected physical behavior is recovered: $\alpha\to 1$ (i.e., the spacetime tends towards flatness) in dispersive regions, while $\alpha\to 0$ precisely where the coordinate singularity forms. This is mirrored in the scalar field, which exhibits sharply localized and strongly negative values in $r$ when $C\to 1$. These limiting behaviors are captured thanks to the boundary conditions~(\ref{eq:BC_conditions}) which enforce the correct asymptotics at distances sufficiently far from the source of perturbation.

\section{Conclusions}
\label{sec:Conclusions}

The gravitational collapse of a massless scalar field remains a demanding benchmark for numerical methods in numerical relativity, as it exhibits critical behavior at the boundary between dispersion and black hole formation. The original investigations conducted by Choptuik~\cite{Choptuik_seminal,Choptuik_1998} tackled the solution of the field equations using finite-difference methods. Those equations govern the evolution of a massless scalar field $\phi$ in GR, minimally coupled to the spacetime metric $g_{\mu\nu}$, and, provided that the initial data is sufficiently strong, black hole formation can occur in the model.  
The studies of~\cite{Choptuik_seminal,Choptuik_1998} probed the existence of a critical value, $p^{*}$, for the control parameter fully describing the initial scalar field distribution. As $p\to p^{*}$, the mass of the black hole, $M_{\mathrm{BH}}$, scales with the deviation $\left|p-p^{*}\right|$ via a universal exponent $\gamma$. Additional phenomena, such as self-similarity and solution echoing, were also observed.

The work reported in this paper has revisited this challenging physical problem, focusing on solutions at and around the critical point. Our primary aim has been to demonstrate that neural network-based methodologies, such as Physics-Informed Neural Networks (PINNs), can serve as a viable and useful alternative to finite differences. To this end, we have considered four different initial-value problems for the field, using the amplitude of the initial kernel, $\phi_{0}$, as the control parameter, with the critical case included among them. We have addressed these cases using several recent methodologies from the literature and introduced our own, termed ModPINN, which has been tailored to this class of scenarios, including several novel aspects from recent works such as coordinate embedding, quadratic neural layers, and learnable activation functions. Our model delivers slightly better critical solutions with lower variability than other state-of-the-art approaches, even though all of them have led to very competitive predictions. Furthermore, by implementing a relatively simple algorithm that moves the spacetime sampling points $(t,r)$, we treat them as particles subject to a ``force'' with a prescribed acceleration modulated via the $\Lambda$ hyperparameter in~(\ref{eq:m_density}), which indicates how quickly points migrate to regions with higher physical loss. The key consequence is that we are able to work with comparatively small sets, unlike traditional numerical methods that resolve black holes by brute force---highly densifying the mesh in spacetime regions that are known a priori to be most challenging.

Irrespective of the specific models used, much of the substantive effort resides in the design of the physical loss. The boundary conditions for $r\gg 0$ introduced in~(\ref{eq:BC_conditions}), are formulated in terms of the derivatives of the physical variables, whose enforcement leads to the asymptotic behaviors that these variables are expected to have far from the source of perturbation. These conditions, as well as the mechanisms of temporal and spatial enforcement, have proved to be crucial and, in some sense, help to emulate the adaptive refinements of finite-difference schemes without incurring additional computational effort. In the analysis displayed in figure~\ref{fig:results_1}, we have examined the robustness of ModPINN with respect to several factors, including the density of sampling points  and the width and depth of the network. We have found that wider and deeper architectures indeed facilitate better solutions, and they also perform relatively well on comparatively large spatiotemporal grids such as $256\times 256$ points---still small in comparison with those used in finite-difference schemes. Overall, these configurations exhibit greater robustness and smoother performance across independent trials. While further analysis could be considered, this allows us to highlight a notable advantage of PINNs: aside from hardware constraints, there is substantial flexibility to scale model size and sampling, and the resulting $N_{\mathrm{t}}\times N_{\mathrm{r}}$ grids in deep learning setups are orders of magnitude smaller than those typically required by finite differences.

Across all models considered for the target problems, we can conclude that physics-informed deep-learning approaches are indeed capable of resolving gravitational collapse driven by a massless scalar field in a spherically symmetric, time-dependent spacetime. By judiciously selecting activation functions to restrict variables to their physically feasible ranges and by formulating boundary conditions in derivative form, we obtain solutions close to the ground truth. While there is still room for improvement, the results indicate that these models are progressing in the right direction. Nevertheless, a PINN can be seen as an alternative (or surrogate) to a finite-difference numerical solver. Broadly speaking, PINN-type models employ the governing PDEs as components of a constructed loss function---alongside additional information---to solve the physical problem at hand. Consequently, once the initial scalar field is specified, the full evolution of $\phi$, $\alpha$, and $C$ over $(t,r)$ is determined. Thus the problem becomes deterministic; gravitational collapse may or may not occur depending on that initial configuration. Therefore, this imposes certain limitations when investigating aspects such as the universality of the exponent $\gamma$ in the scaling law of the black hole mass $M_{\mathrm{BH}}$ near criticality, as well as the generality of the proportionality constant $c_{f}$ with respect to the family of the initial field. To examine these aspects---and to study the self-similarity of solutions as $p\to p^{*}$---we would need access to a broad set of initial condition sets and their full corresponding solutions. Therefore, addressing these questions will require a different point of view---one that allows us to traverse families of initial conditions with full flexibility and to generate data with relative ease. In this sense, the present project paves the way for future studies that can probe the aforementioned issues, extend to alternative spacetime metrics and matter models, and tackle more complex scenarios in which finite-difference methods could be either inapplicable or impractical due to prohibitive computational costs.

\section{Acknowledgments}
\label{sec:Acknowledgments}

This work has been partially funded by the Valencian Government grant with Reference Number CIAICO/2024/111; the Spanish Ministry of Economic Affairs and Digital Transformation through the QUANTUM ENIA project call-QuantumSpain project, and the European Union through the Recovery, Transformation and Resilience Plan—NextGenerationEU within the framework of the Digital Spain 2025 Agenda. This work has been also supported by the Spanish Agencia Estatal de Investigación (grant PID2024-159689NB-C21) funded by MCIN/AEI/10.13039/501100011033 and ERDF A way of making Europe, by the Prometeo excellence programme grant CIPROM/2022/49 funded by
the Generalitat Valenciana, and by the European Horizon Europe staff exchange (SE) programme HORIZON-MSCA2021-SE-01 Grant No. NewFunFiCO-101086251. MWC is supported by NSERC. R. RdA is supported by PID2020-113644GB-I00 from the Spanish Ministerio de Ciencia e Innovación and by PROMETEO/2022/69 funded by the Generalitat Valenciana.


\printbibliography


\end{document}